\documentclass[aps,amsmath,amssymb,prd,showpacs,floatfix,preprint,superscriptaddress,nofootinbib,12pt]{article}
\pdfoutput=1

\usepackage{jheppub}
\usepackage{slashed,mathrsfs}
\usepackage{amsmath,epsfig}
\usepackage{amssymb,amsfonts}
\usepackage{latexsym}
\usepackage[latin1]{inputenc}
\usepackage{subeqnarray}
\usepackage{xcolor}

\usepackage{graphicx}
\usepackage{ulem}
\usepackage{float}
\usepackage{subfig}

\def \beq{\begin{equation}}
\def \eeq{\end{equation}}
\def \bse{\begin{subequations}}
\def \ese{\end{subequations}}
\def \bea{\begin{eqnarray}}
\def \eea{\end{eqnarray}}
\def \bem{\begin{displaymath}}
\def \eem{\end{displaymath}}
\def \bem{\begin{pmatrix}}
\def \eem{\end{pmatrix}}

\newcommand{\mymu}[1]{\mu^{\phantom{\dag}}_{\text{#1}} }
\newcommand{\ads}[1]{\text{AdS}$_{#1}$}
\newcommand{\mypar}{{\mkern3mu\vphantom{\perp}\vrule depth 0pt\mkern3mu\vrule depth 0pt\mkern3mu}}

\allowdisplaybreaks[3]
\title{ \center \Huge
Conjecture on the Butterfly Velocity across a Quantum Phase Transition
}

\author[\dag]{Matteo Baggioli}
\affiliation[\dag]{Crete Center for Theoretical Physics, Institute for Theoretical and Computational Physics\\ Department of Physics, University of Crete, 71003
Heraklion, Greece.}
\author[\star]{, Bikash Padhi}
\author[\star]{, Philip W. Phillips}
\author[\star]{, Chandan Setty}
\affiliation[\star]{Department of Physics and Institute for Condensed Matter Theory,
University of Illinois, 1110 W. Green Street, Urbana, IL 61801}

\emailAdd{mbaggioli@physics.uoc.gr}
\emailAdd{bpadhi2@illinois.edu}
\emailAdd{dimer@illinois.edu}
\emailAdd{csetty@illinois.edu}

\abstract{
We study an anisotropic holographic bottom-up model displaying a quantum phase transition (QPT) between a topologically trivial insulator and a non-trivial Weyl semimetal phase. We analyze the properties of quantum chaos in the quantum critical region. We do not find any universal property of the Butterfly velocity across the QPT. In particular it turns out to be either maximized or minimized at the quantum critical point depending on the direction of propagation. We observe that instead of the butterfly velocity, it is the dimensionless information screening length that is always maximized at a quantum critical point. We argue that the null-energy condition (NEC) is the underlying reason for the upper bound, which now is just a simple combination of the number of spatial dimensions and the anisotropic scaling parameter. 
}
\preprint{ITCP-IPP 2018/23, CCTP-2018-4}

\begin{document}
\maketitle

\renewcommand*{\thefootnote}{\arabic{footnote}}	
\setcounter{footnote}{0}


\section{Introduction}
\begin{figure}
\center
\includegraphics[width=12cm]{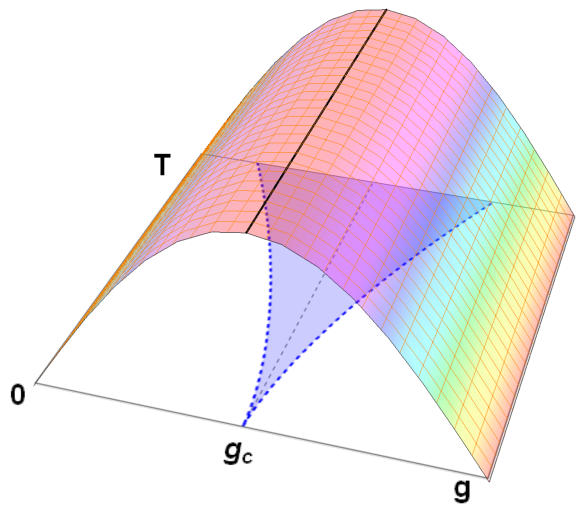}
\caption{An intuitive sketch of the results of \cite{ShenQCP} regarding the behavior of the Liapunov exponent $\lambda_L$ on the quantum critical region (blue region). The exponent $\lambda_L$ gets its maximum value (black line) at the quantum critical point $g=g_c$. It is not clear, and indeed the purpose of our investigation, if the butterfly velocity $v_B$ display a similar behaviour or not.}
\label{figsketch}
\end{figure}

One of the remarkable claims that has arisen in recent years from the unexpected connection between quantum chaos, quantum criticality, transport and universality is that a many-body system exhibiting a quantum phase transition the Lyapunov exponent is maximized at the critical point \cite{ShenQCP}, and the butterfly velocity shows some characteristic behavior across this point \cite{LingVbQCP}. The Lyapunov exponent determines the (late time) growth of out-of-time correlation (OTOC) function,
\beq
\langle\,\left[\mathcal{V}(\vec{x},t)\, , \,\mathcal{W}(0,0)\right]^2\,\rangle_\beta\,\sim\,e^{\lambda_L\,\left(t-t^*\,-\,|\vec{x}|/v_B\right)},\label{OTOCdef}
\eeq
where $\mathcal{V},\mathcal{W}$ are two local Hermitian operators, $\lambda_L$ the Lyapunov exponent, $t^*$ is the so called scrambling time and $\beta$ is just the thermal timescale. 
The appearance of the butterfly velocity in this correlation function motivates it as the relevant velocity for defining bounded transport \cite{BlakePRD16}. The monotonic growth of the Lyapunov exponent at a quantum critical point and its subsequent decrease away from the critical point determined by some non-thermal coupling constant, $g$ is sketched in figure~\ref{figsketch}.  See also \cite{Ling:2016wuy} for an exploration of the connection between quantum chaos and thermal phase transitions. Moreover, this behavior is believed to hold also at finite but low temperature inside the quantum critical region. Preliminary studies connected to the proposal of \cite{ShenQCP} and to the onset of quantum chaos across a QPT have been already performed within the holographic bottom-up framework in \cite{LingVbQCP}. Nevertheless a complete study, beyond simple models, is still lacking. A full analysis of this problem appears to be in order in view of the recent experiments where the OTOC has been measured using Lochsmidt echo sequences \cite{ExpColdAtoms} and NMR techniques \cite{ExpNMR}.

The aim of this paper is indeed to understand the onset of quantum chaos across a quantum phase transition in more complicated holographic models displaying a quantum phase transition. In particular, we will perform our computations in the holographic bottom-up model introduced in \cite{Landsteiner:2015lsa, Landsteiner:2015pdh} which exhibits a QPT between a trivial insulating state and a Weyl semimetal. The particular new wrinkle we bring to bear on the transition is the presence of anisotropy.  In other words, the rotational group $SO(3)$ is broken to the $SO(2)$ subgroup by an explicit source in the theory. As a consequence of the underlying anisotropy, we can define two butterfly velocities that we will denote as $v_\mypar$ and $v_\perp$, where $\mypar$ denotes the direction(s) of the anisotropy, while $\perp$ in the direction perpendicular to it. Throughout the paper, we use this notation for all such directional quantities.

The results of our paper show that while the perpendicular velocity $v_\perp$ displays a behavior similar to that in \cite{ShenQCP}, the parallel one $v_\mypar$ does not. In particular, the butterfly velocity along the anisotropic direction will not display a maximum at the critical point $g=g_c$ but rather a \textit{minimum}. We pinpoint as the origin of this violation, the presence of anisotropy itself\footnote{Effects of anisotropy on the butterfly velocity were previously investigated in \cite{Blake:2017qgd,Ahn:2017kvc,Viktor,Dimitrios1}.}. Interestingly, the bound on the viscosity is also violated in an anisotropic system \cite{KSSviolationAniso,Jain:2015txa} and by a strong magnetic field \cite{Finazzo:2016mhm,Critelli:2014kra}. Here the mechanism leading to the violation of the bound are very analogous, that is \textit{explicit} breaking of the $SO(3)$, leading to spatial-anisotropy. Note that  \cite{Erdmenger:2012zu} spontaneous breaking of rotation symmetry, despite leading to a non-universal value for $\eta/s$, does not provide a violation of the KSS bound. We expect this to be case for butterfly velocity as well. Here $\eta$ is the viscosity and $s$ is the entropy density. 

To understand if any universal statement can be made about the butterfly velocity, especially in the presence of anisotropy, we identify a quantity related to the spatial-spread of information which is insensitive to the breaking of the $SO(D)$ symmetry, where $D$ is the number of space dimensions. We do so by computing the OTOC holographically. Given an anisotropic bulk spacetime of the form
\begin{equation}
ds^2\,= -\,\textsl{g}_{tt}(r)\,dt^2\,+\,\textsl{g}_{rr}(r)\,dr^2\,+\,h_{\perp}(r)\,d\vec{x}_\perp^2\,+\,\,+\,h_{\mypar}(r)\,d\vec{x}_\mypar^2,
\end{equation}
where we denote by $\mypar$ the $D_{\mypar}$ anisotropic directions and with $\perp$ the $D_{\perp}$ remaining directions. The butterfly velocities can be computed for this background as ($\eta = \perp, \mypar$) ${v}_{\eta}\,= {\lambda_L}/{M_{\eta}},$ where $\lambda_L=2\pi/\beta$ is the Lyapunov exponent and all the quantities are computed at the horizon. The parameter $M_{(\perp,\mypar)}$ controls the screening of the information spreading in the $(\perp,\mypar)$ directions,
\begin{equation}
\psi(t,x_\eta)\sim\,e^{\lambda_L\,t\,-\,M_\eta \,|x_i|},
\end{equation}
and it clearly depends on the warp factor $h_\eta$. As a consequence, the butterfly velocity can not represent a good and universal quantity in the presence of anisotropy. Contrastly, we can define a dimensionless quantity controlling the screening of information through
\begin{equation}
\mu^2\,\equiv\,\frac{M_\eta^2}{h_\eta (r_0)}\, .
\end{equation}
The factor $h_\eta(r_0)$ is indeed the reason why we see dissimilar result from \cite{ShenQCP}  in an anisotropic setup. The important point is that our new physical parameter $\mu$ has no spatial dependence and hence, it is completely insensitive to any anisotropy present in the system. Our proposal is to consider the dimensionless information screening length $\mathrm{L}$, which can be defined as $ \mathrm{L}  \equiv 1/{\mu}$. Our claim can be rephrased as \textit{the dimensionless information screening length $\mathrm{L}$, which can be  defined via the OTOC, is always maximum at the quantum critical point}. Moreover, for a theory passing through a Lifshitz-like critical point, given the number of spatial directions $D_\perp$ which scales similarly as time and the number of the directions $D_\mypar$ which has an anisotropic scaling, $\beta_0$, the conjecture regarding $\mathrm{L}$ can be restated as
\begin{equation}
2\mathrm{L}\,\leq\,\frac{1}{D_\perp\,+\,\beta_0\,D_\mypar}\, .
\label{def3}
\end{equation}
We will later see that such a bound can be justified from NEC and in our model this is saturated at the quantum critical point $g=g_c$. In a similar spirit, \cite{Isoperimentry17} points out a bound on the butterfly velocity for an isotropic space with different warp factors appearing along the $r$, $t$ directions, $\textsl{g}_{tt}(r), \textsl{g}_{rr}(r)$. Since in our case $\textsl{g}_{tt}(r)  \textsl{g}_{rr}(r) = 1$, we always saturate their bound. 

The paper is organized as follows. In section \ref{sec:HoloModel}, we present the holographic model and its main, and known, transport features.  In section \ref{sec:chaos}, we study the onset of quantum chaos in the model and in particular the butterfly velocity and the related conjectured bound.  Conclusions are reached in \ref{disc}. In appendices \ref{app1} and \ref{vbAPP} we provide more  technical details about our computations.


\section{The Holographic model}
\label{sec:HoloModel}
We begin by reviewing the holographic model of \cite{Landsteiner:2015lsa, Landsteiner:2015pdh} which exhibits a QPT from a topologically non-trivial Weyl semimetal to a trivial insulating phase. Although the boundary theory exhibiting this topological transition in Eq.  \eqref{eq:DiracLag} is a free theory, the holographic bulk theory strictly describes a strongly correlated system. The hope is that they share the same set of symmetries, thereby capturing the essential properties of the phase transition, if not all the details of the transport pertaining to interacting physics. Note that this is a phase transition in a certain topological invariant (such as  Chern number) and not in the symmetries; thus, one can not probe it through the free energy density as it never depends on any topological term in the action. The order parameter is represented by the anomalous Hall conductance, $\sigma_{\text{AHE}}$ which is zero in the trivial gapped phase and finite in the Weyl semimetal phase.

\begin{figure}
\center
\includegraphics[width=0.88 \columnwidth , height= 0.4 \columnwidth]{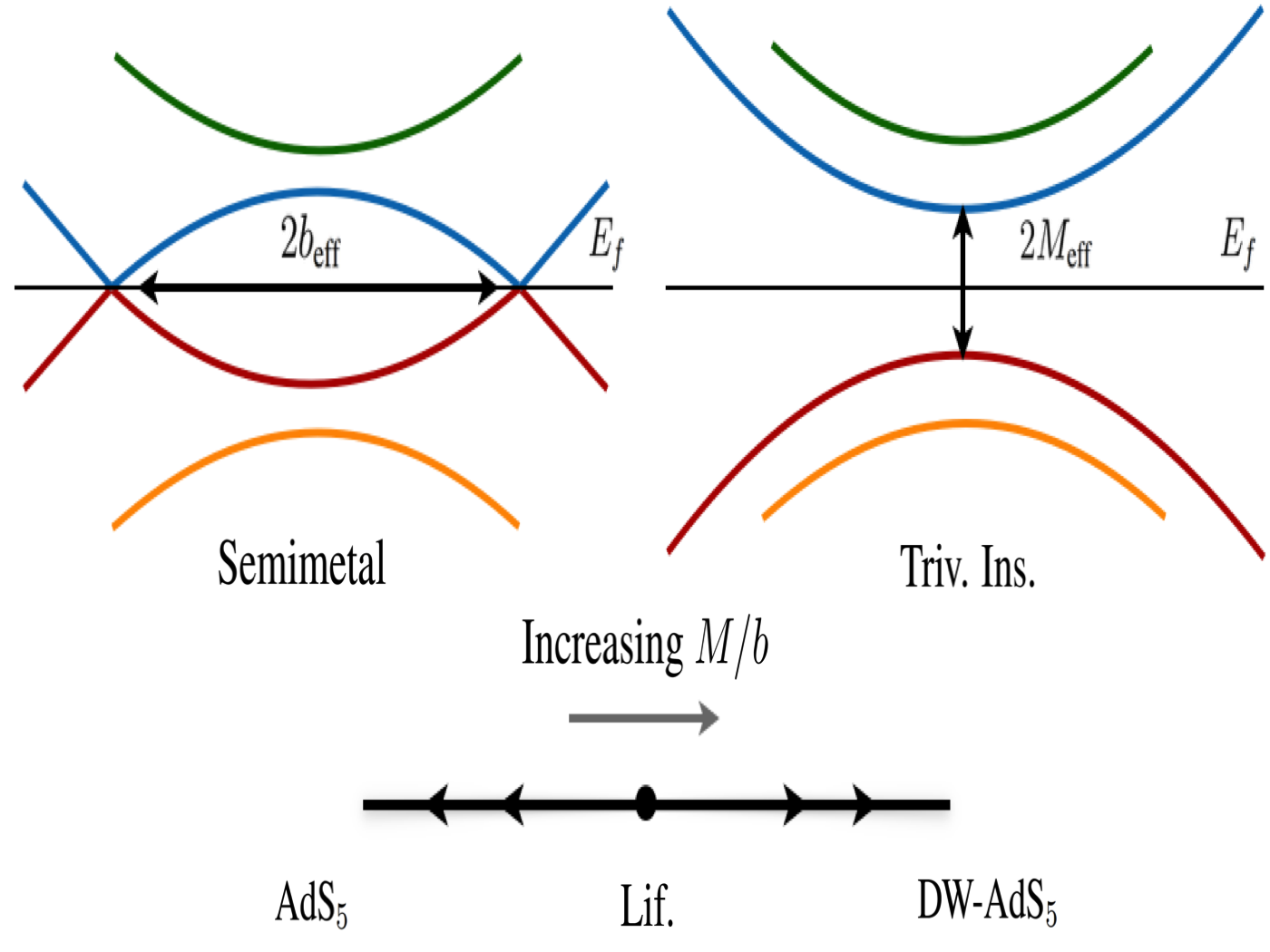}
\caption{There is a topologically non-trivial to trivial phase transition as $M/b$ increases (from left to right). Due to broken time-reversal symmetry, the 2-fold degeneracy has been lifted, giving rise to 4-bands in Eq. \eqref{eq:DiracLag}. The (top) left figure has two Weyl nodes (a pair of Dirac cones where the bands cross) separated by $2\vec{b}_{\text{eff}}$ in momentum. The band structure on the (top) right is that of a topologically trivial insulator with an explicit band-gap $2M_{\text{eff}}$. At the QCP, the two Dirac cones merge together, giving rise to a Lifshitz fixed point (black dot in the bottom figure) with a scaling anisotropy along the same direction as $\vec{b}$. In the holographic picture, away from the QCP, the theory flows to two different types of (deep IR) near-horizon geometries, \ads{5} (Weyl semimetal phase) or domain wall-\ads{5} (trivial insulator phase). The figure shares some resemblances with those in \citep{Landsteiner:2015lsa,TopDownWeyl}.}
\label{figsketch2}
\end{figure}

\subsection{Weyl Semimetals}
Weyl semimetals are a class of three dimensional topological materials characterized by (point) singularities in the Brillouin zone (BZ) at which the band gap is zero. This peculiar property gives rise to exotic transport phenomena ( see \cite{QiWeylRev} for a comprehensive review). Quasiparticle excitations near such  band-touching points, also called Weyl nodes, can be described by (left- or right-handed) Weyl spinors. In a time-reversal symmetry broken insulator, the left- and right- Weyl nodes are separated in the BZ which can be controlled by a chiral or axial gauge potential, $\vec{b}$. It is the interplay of this axial field and the (chiral) mass of the spinor, $M$, that gives rise to different phases (see figure \ref{figsketch}). Deep in the semi-metal phase, $b \,\,(\equiv |\vec{b}|)$ is much larger and $M$ simply renormalizes it causing a reduced node-separation equal to \cite{BalentsBurkov} $b_{\text{eff}} = b\, (1 - \bar{M}^2)^{1/2}$, where $\bar{M}\equiv M/b$. On the other hand, for a larger $M$, renormalization by a weaker $b$ reduces the gap to $M_{\text{eff}} = b \, (\bar{M}^2 - 1)^{1/2}$. Thus, the semimetal-insulator phase transition occurs at $\bar{M}_c \sim \mathcal{O}(1)$. The continuum description capturing this physics is \cite{PallabWeylQFT}
\bea
\mathcal{L} = \bar{\psi}\, (i \slashed{\partial} - e \slashed{B} - \gamma^{\phantom{\dag}}_5 \vec{\gamma} \cdot \vec{b} + M )\, \psi \, .
\label{eq:DiracLag}
\eea
Here the slash denotes contraction by Dirac gamma matrices, $\gamma_\mu$. The matrix $\gamma_5 = i \gamma_0 \gamma_1 \gamma_2 \gamma_3$ allows one to project the Dirac spinors, $\psi$, into the chiral sectors, $ \psi_{L,R} = (1 \pm \gamma^{\phantom{\dag}}_5 ) \psi$.  $B_\mu$ is the electromagnetic gauge potential; without loss of generality \cite{BurkovPRB}, we choose the axial gauge potential to be $\vec{b} = b \,\hat{e}_z$. The axial symmetry, however, is anomalous ($dJ_5 \neq 0$), leading to a non-conservation of the number of particles of given chirality. This can be seen \cite{XuAHEPRL} in the response of the axial current, $\vec{J}_5 \sim \vec{b}_{\text{eff}} \times d\vec{A}$, that is the anomalous Hall conductance, $\sigma_{\text{AHE}} \sim {b}_{\text{eff}}$. This clearly vanishes in the insulating phase, that is for sufficiently large $M$. The mass term and the axial term act as relevant deformations. Thus, with increasing $\bar{M}$, the theory moves from UV to IR thereby traversing through a fixed point at $\bar{M}_c$.

\subsection{Holographic Weyl Semimetal}
Now we turn to the holographic model of the above phase transition. The bulk action takes the form (fixing $2 G_N^2=L=1$, where $G_N$ is Newton's constant, and $L$ the AdS radius):
\begin{align}
\mathcal{S}\,=\,\int d^5x\,\sqrt{-g}\,\Big[\phantom{]}&R\,+\,12\,-\,\frac{1}{4}\,F^2\,-\,\frac{1}{4}\,F_5^2 \,+\,\frac{\alpha}{3}\epsilon^{\mu\nu\rho\sigma\tau}\,A_\nu\,\Big({F^5}_{\nu\rho}\,{F^5}_{\sigma\tau}\,+\,3\,F_{\nu\rho}\,F_{\sigma\tau}\Big)\,\nonumber\\
&-\left(D_\mu \Phi\right)^{*}\,\left(D_\mu \Phi\right)\,-\,V\left(\Phi\right)\phantom{[}\Big]\label{action}
\end{align}
The bulk fields are an electromagnetic vector $U(1)$ gauge field $B_\mu$ with fields strength $F=dB$, an axial gauge field $A_\mu$ with field strength $F_5=dA$ and a complex scalar field $\Phi$ charged under the axial $U(1)$ symmetry. The covariant derivative is defined as $D_\mu \Phi=\partial_\mu -i q A_\mu \Phi$, and the scalar potential is chosen to be $V(\Phi)=m^2|\Phi|^2+\frac{\lambda}{2}|\Phi|^4$. Since the phase of the scalar field is not a dynamical variable, with out loss of generality we assume it to be real. The mass of the field, $m = \sqrt{\Delta(\Delta-d)}$ controls the scaling dimension, $\Delta$, of the boundary operator corresponding to $\Phi$. Throughout the paper, we will use $d$ as the space-time dimension of the boundary field theory, occasionally denoting the boundary spatial dimension as $D=d-1$. From the mass deformation in Eq. \eqref{eq:DiracLag} and the above relation, it is clear that one needs to choose $m^2=-3$ (see \cite{TopDownWeyl} for different choices of $m^2$ and \cite{Ammon:2016mwa,Ammon:2018wzb} for further studies of the model), such that the dual operator has conformal dimension $\Delta=3$. Note that this imaginary mass is perfectly allowed within AdS/CFT since it is with in the Breitenlohner-Freedman (BF) bound, $m^2 \geq -d^2/4$. The UV boundary conditions for the vector and scalar field are chosen to be 
\beq
\lim_{r \rightarrow \infty} \, r \,\Phi = M \quad , \quad \lim_{r \rightarrow \infty} \, A_z = b\, ,
\eeq
where both $M$ and $b$ represent a source for the corresponding dual operators. The parameter $b$ can be thought as an axial magnetic field that explicitly breaks the rotational $SO(3)$ symmetry of the boundary to the $SO(2)$ subgroup. From figure \ref{figsketch2}, one can see that this controls the effective separation between Weyl nodes. On the contrary, the source $M$ for the scalar field is simply introducing the mass scale required by the physics of the problem. Note the presence of two more (bulk) free parameters in the problem; the quartic coupling, $\lambda$, controls the location of the quantum critical point (QCP) by changing the depth of the effective potential of $\Phi$, and the charge $q$ relates to  the mixing between the operators dual to $\Phi$ and $A_\mu$. Following \cite{Landsteiner:2015lsa, Landsteiner:2015pdh} we fix these parameters to $q=1$, and $\lambda=1/10$, which fixes $\bar{M}_c$ to $0.744$. The generic solution of the system is given by the following ansatz
\begin{align}
&ds^2 = - f(r) dt^2 + \frac{dr^2}{f(r)} + g(r) \left( dx^2 + dy^2 \right) + h(r) \,dz^2 \, ,\\
&A\,=\,A_z(r)\,dz\,,\quad \Phi=\phi(r).
\label{ansatz}
\end{align}

Although not necessary for computing the butterfly velocity, we will first discuss the behavior of zero-temperature solutions for understanding the various low-temperature limits. For finite temperature, we assume the presence of a black hole horizon at $r=r_0$ such that $f(r_0)=0$. For the zero temperature background there is a Poincar\'e horizon at $r_0 = 0$, and $f(r) = g(r)$. There are tree types of solutions at zero temperature -- (i)  insulating background (for $\bar{M} > \bar{M}_c$), (ii) critical background (for $\bar{M} = \bar{M}_c$), and (iii) semimetal background (for $\bar{M} < \bar{M}_c$). These solutions can be obtained by solving the equations of motion, the details of which we discuss in the Appendix \ref{app1}. We quote the results here (up to leading order near the IR).\\[0.2cm]
\textit{Insulating background}. --- Similar to a zero-temperature superconductor, the near-horizon geometry of a topologically trivial insulator is an \ads{5} domain-wall 
\bea
f(r) = (1 + \frac{3}{8 \lambda})  r^2 \, , \quad h(r) = r^{2} \,, \quad A_z(r) = a_1 r^{\beta_1} \, , \quad \phi(r) = \sqrt{\frac{3}{\lambda}} + \phi_1 r^{\beta_2} \,.
\label{bkg:Ins}
\eea
Here $a_1$ is fixed to $1$ and $\phi_1$ is treated as a shooting parameter. Exponents $\beta_{1,2}$ can be expressed as functions of $(m,\lambda,q)$, and are $(2.69, 0.29)$ for our choice of parameters. Thus, the near-horizon value of $A_z$ is always zero, and that of $\phi$ is $\sqrt{3/\lambda}\,$ (for $\lambda = 1/10$, it is $\phi(r_0)\simeq 5.477$).\\[0.2cm]
\textit{Critical background}. --- This solution is \textit{exact} and displays an anistropic Lifshitz-like scaling parametrized by $\beta_0$,
\bea
f(r) = f_0 r^2 \, , \quad h(r) = h_0 r^{2\beta_0} \,, \quad A_z(r) = r^{\beta_0} \, , \quad \phi(r) = \phi_0\,.
\label{bkg:Lifs}
\eea
The scaling anisotropy is \textit{explicitely} induced by the source of the axial gauge field $A_\mu$, hence is along the direction of $\vec{b}$. The parameters $(f_0, h_0, \beta_0, \phi_0)$ are determined by fixing $(m,\lambda,q)$. For the parameter choice mentioned previously, we have $(f_0, h_0, \beta_0, \phi_0) \simeq (1.468, 0.344, 0.407, 0.947)$. 
From the zero-temperature equations of motion, it can be shown that $\beta_0 = -2q^2/(m^2 + \lambda \phi_0^2 - 2q^2)$ and is always $\leq 1$ owing to the NEC, and regularity of solutions demands $\beta_0 > 0$  \cite{TopDownWeyl}. Thus, the near-horizon value of $A_z$ at criticality is always zero, whereas that of $\phi$ is $\phi_0$.\\[0.2cm]
\textit{Semimetal background}. --- The following solution describes the near-horizon geometry of the semimetal phase, which is simply \ads{5} 
\bea
f(r) = r^2 = h(r)\,,\,\, A_z(r) = a_1 + \frac{\pi a_1^2 \phi_1^2}{16 r}\, e^{-\frac{2 a_1 q}{r}}\, , \,\, \phi(r) = \sqrt{\pi} \phi_1 \left( \frac{a_1 \phi_1}{2 r} \right)^{3/2} e^{- \frac{a_1 q}{r}} . 
\label{bkg:Weyl}
\eea
The $\lambda$ dependence is hidden in higher order terms.  Note in this case, the near horizon solution of $A_z$ is finite; $a_1$, however, $\phi(r_0)$ vanishes. Figure \ref{figbulk} and \ref{fignum} of Appendix \ref{app1} provide the full $A(r)$ and $\phi(r)$ functions for various values of $\bar{M}$. The apparent deviations of $A(r_0)$ and $\phi(r_0)$ from the IR asymptotes described above owes to the fact that we obtain the solutions for a small but finite temperature up to order $\mathcal{O}(\bar{T})$, where $\bar{T} \equiv T/b$. We will treat $\bar{M}$ and $\bar{T}$ as the free parameters in the theory to control the phase transition.

\subsection{Anomalous Transport}
\begin{figure}[htp]
\centering
\includegraphics[width=0.76 \columnwidth , height= 0.6 \columnwidth]{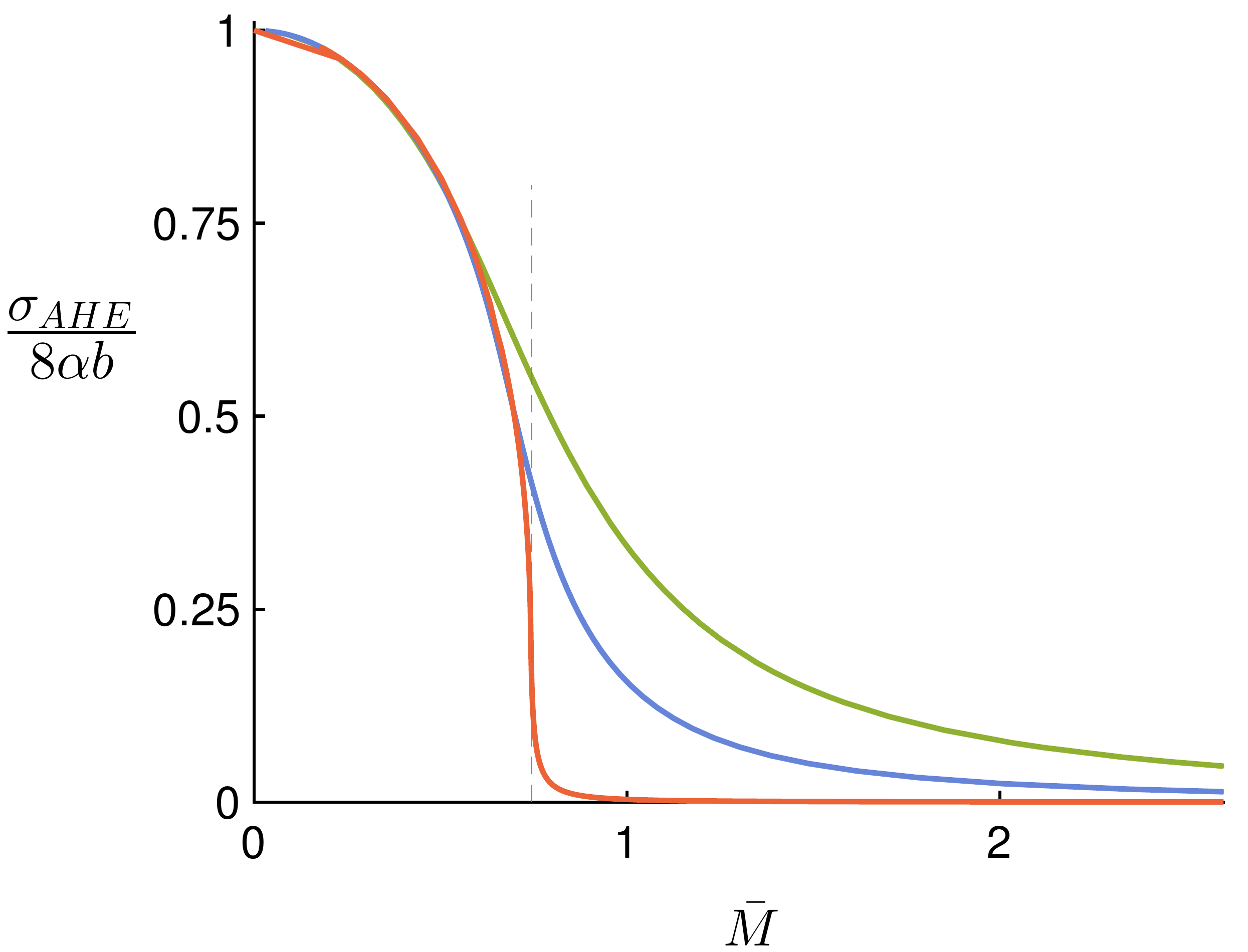}
\caption{Anomalous Hall conductivity (obtained from $\sigma_{xy}$) as a function of the dimensionless mass parameter $\bar{M}$ for temperatures $\bar{T}=0.1,0.05,0.005$ (from green to orange). Note for a very low temperature the conductivity sharply drops to zero at a critical value, $\bar{M}_c \sim 0.74$. This marks the semimetal-insulator topological phase transition.}
\label{fig:Cond}
\end{figure}
\begin{figure}[htp]
\centering
\includegraphics[width=0.48 \columnwidth , height= 0.42 \columnwidth]{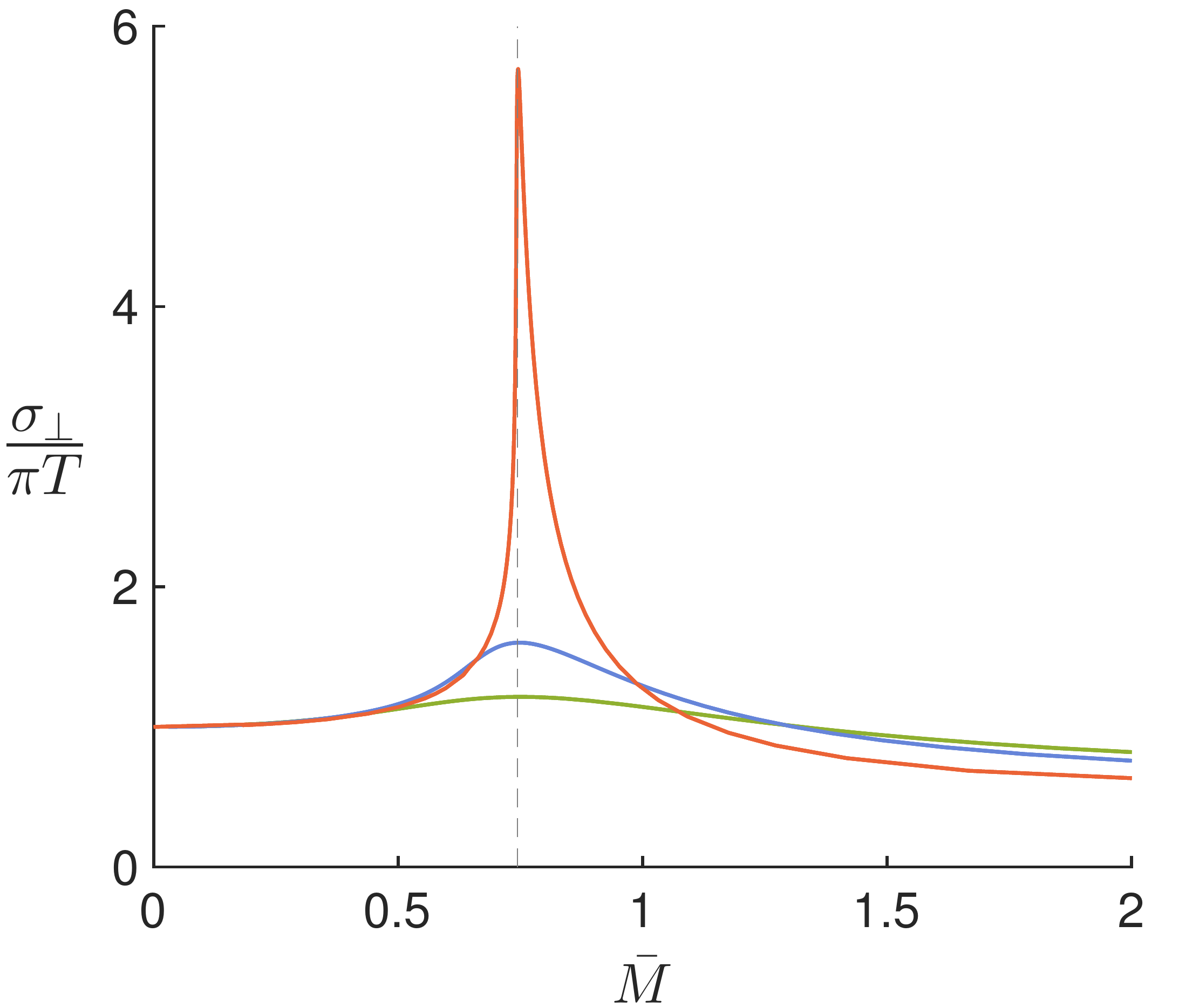}
\includegraphics[width=0.48 \columnwidth , height= 0.42 \columnwidth]{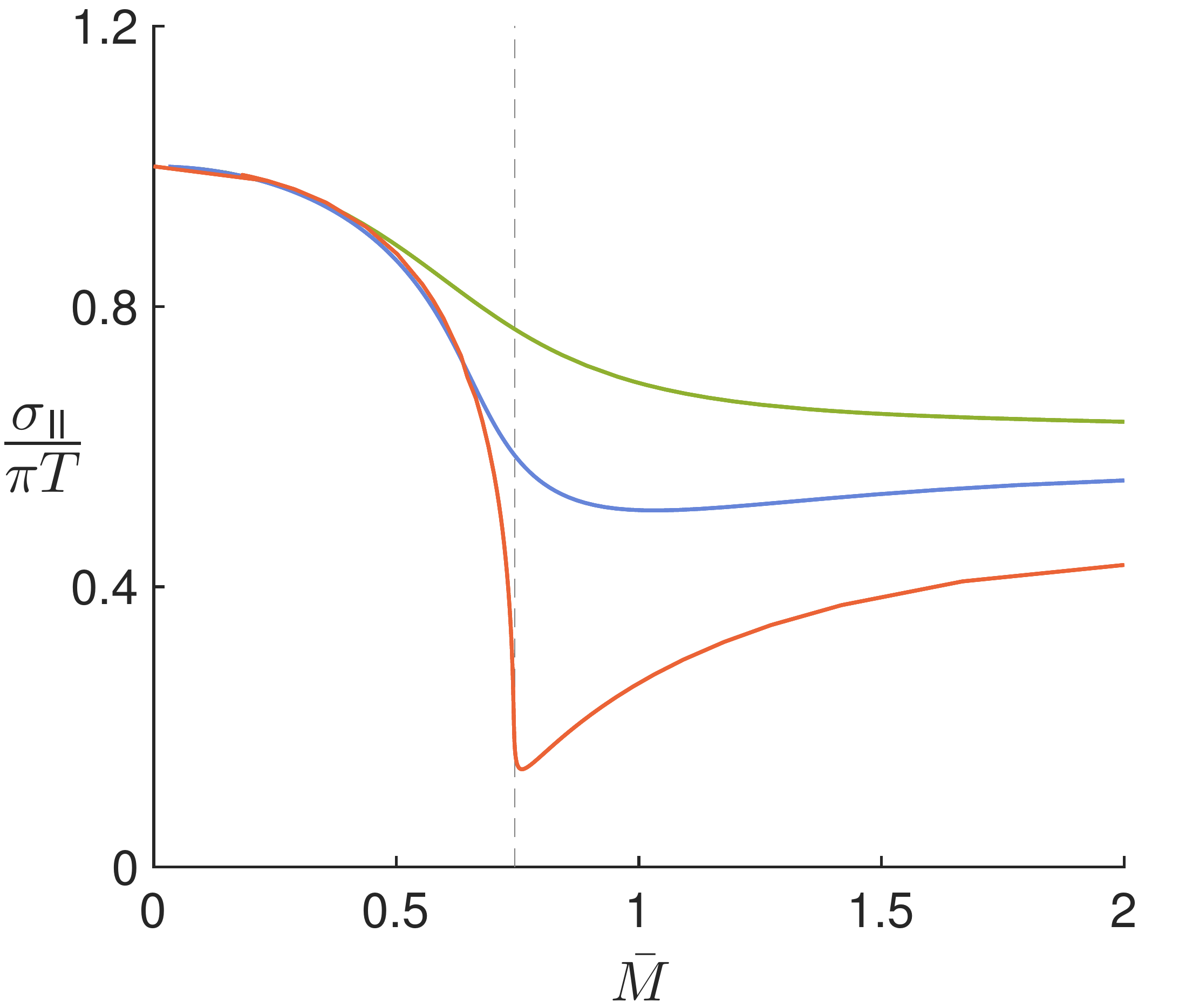}
\caption{Longitudinal conductivities $\sigma_{xx},\,\sigma_{zz}$ in function of the dimensionless quantum parameter $\bar{M}$ for temperatures $\bar{T}=0.1,0.05,0.005$ (from green to orange).}
\label{fig2}
\end{figure}

As mentioned before, the order parameter for the QPT is the anomalous Hall conductivity. The DC,  limit of all the conductivities  can be extracted from (for both zero and finite temperatures) horizon data as follows
\beq
\sigma_{\text{AHE}}=8 \,\alpha \,A_z(r_0)\, , \quad \sigma_{\perp}=\sqrt{h(r_0)}\,,\quad \sigma_{\mypar}=\frac{g(r_0)}{\sqrt{h(r_0)}}.
\label{eq:CondForm}
\eeq
Here $\sigma_\mypar$ is just a short hand for $\sigma_{zz}$ and '$\perp$' refers to the conductivity matrix elements, $\sigma_{xx},\, \sigma_{yy}$, and should not be mistaken for the transverse conductivity. In figure  \ref{fig:Cond} and \ref{fig2}, we plot the above conductivities as functions of $\bar{M}$, for various temperatures $\bar{T}$. We discuss them individually, starting from their zero-temperature behavior. In order not to sacrifice numerical stability, we confine our lowest temperature value to $\bar{T}=0.005$ and treat it as zero temperature.

\begin{figure}[htpb]
\center
\subfloat{\includegraphics[width=0.46 \columnwidth , height= 0.4 \columnwidth]{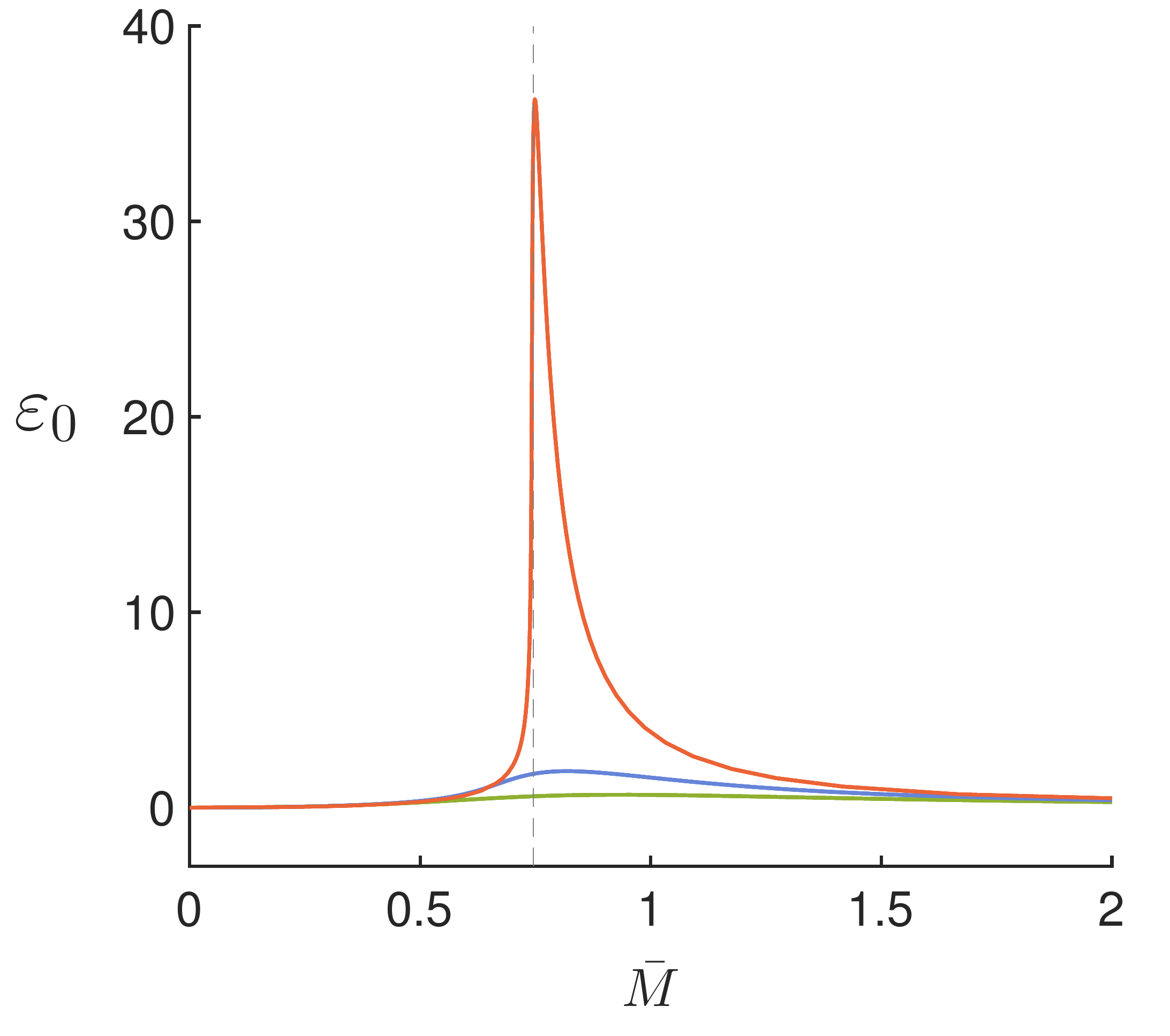} 
 \label{figAN}}
\subfloat{\includegraphics[width=0.45 \columnwidth , height= 0.4 \columnwidth]{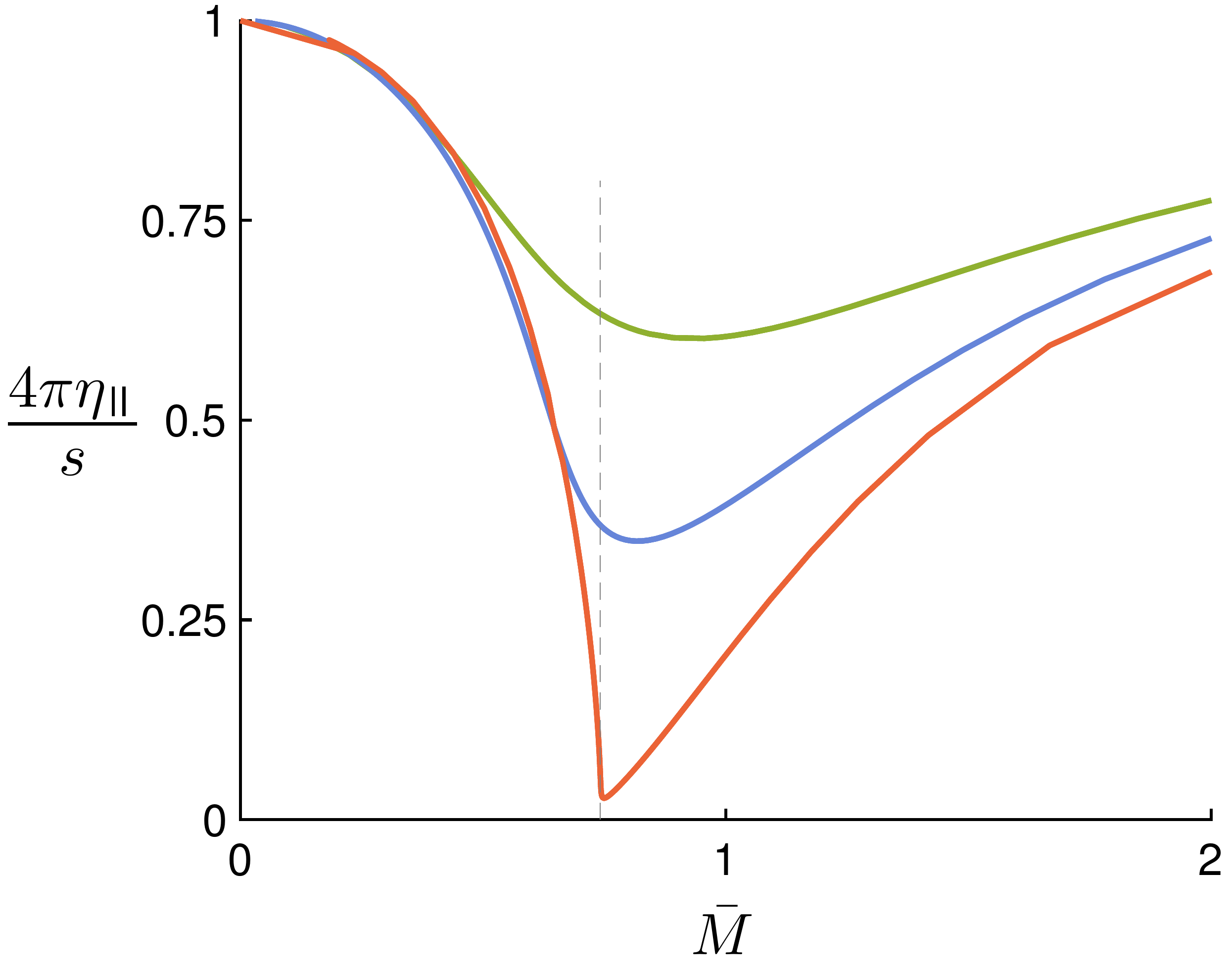}
\label{figeta} }
\caption{(Left) The anisotropy parameter $\varepsilon_0$ evaluated at the horizon for various $\bar{T}=0.1,0.05,0.005$ (from green to orange). (Right) Viscosity to entropy ratio, $4\pi \eta_\parallel /s$ along the anisotropic direction for $\bar{T}=0.005,0.05,0.1$. The viscosity is given in terms of the horizon data as $\eta_\parallel=g^2(r_0)/\sqrt{h(r_0)}$ \cite{LandsteinerViscosity}. The violation of the KSS bound is evident. On the contrary the ratio along the isotropic direction saturates exactly the KSS bound $1/4\pi$ and it is not shown here.
}
\end{figure}

Note that $\sigma_{\text{AHE}} \sim A_z(r_0)$, and from the discussion of the zero-temperature solutions, we see  $\sigma_{\text{AHE}}$ is finite only for $\bar{M} < \bar{M}_c$.  A more physical picture could be that since in the IR, the axial gauge field is completely screened \cite{TBDWeyl15}, there are no degrees of freedom that could be coupled to it and hence, it can not be probed any further. As the temperature is increased, the sharp phase transition slowly becomes a cross-over. At zero-temperature, the onset of the semimetal phase is well fitted by $\sigma_{\text{AHE}}\,\propto\,\left(\bar{M}_c\,-\,\bar{M}\,\right)^{\,0.21}$. 
For $M=0$ (or, $\bar{M}=0$), the near-horizon geometry is the deformed \ads{5} background of Eq. \eqref{bkg:Weyl}. With our choice of normalizations, for low temperatures,  $g(r_0)=h(r_0) = \pi^2 T^2$ and hence, $\sigma_{\text{diag}} \simeq \pi T$, which clearly vanishes at $T=0$. The subscript 'diag' collectively refers to all the diagonal components of the conductivity matrix, $\sigma_{xx}, \sigma_{yy}, \sigma_{zz}$. There are two features of $\sigma_{\text{diag}}$ of interest. First, for vanishing $b$ (or, $\bar{M} \gg 1$) the near-horizon geometry is the domain-wall \ads{5} geometry of Eq. \eqref{bkg:Ins}, which makes $\sigma_{\text{diag}} \simeq c \, \pi T$, where $c < 1$ and independent of temperature. This is due to the fact that it is a phase transition between a semimetal-insulator transition and some degrees of freedom are now gapped out in the trivial phase. The reason why the conductivity is still finite in the insulating phase can be understood by computing the ratio of the gapped to un-gapped degrees of freedom \cite{TopDownWeyl}, which eventually becomes a statement about the geometry or more precisely about the holographic a-theorem \cite{RobMyers10}. 
This ratio can be made to vanish by controlling $m^2$ and $\lambda$. Second, and the most relevant for our discussion, is the fact that at the critical point, there are strong divergences at zero temperature. This can be attributed to the anisotropy of the critical point.  For convenience, we define the ratio $\varepsilon_0$ at the horizon (also see figure \ref{figAN}), 
\begin{equation}
\varepsilon_0 \equiv \frac{h(r_0)}{g(r_0)}\,-\,1
\end{equation}
as the measure of spatial anisotropy along the $z$ direction at the horizon. More precisely, from the expressions of the $\sigma_{\text{diag}}$ in Eq.~\eqref{eq:CondForm}, one can see that the ratio of the two at zero temperature becomes
\beq
\frac{\sigma_{\perp}}{\sigma_\mypar} = \frac{h(r_0)}{g(r_0)} \sim r_0^{2(\beta_0-1)},
\label{blow}
\eeq
which clearly diverges at the quantum critical point $\bar{M}_c$. Another way of achieving the same conclusion is to analyze the AC conductivities \cite{Grignani:2016wyz}. From there, or simply from Eq.~\eqref{blow}, we can indeed conclude that $\sigma_{\perp} / \sigma_{\mypar}  \sim \omega^{2(\beta_0-1)}$, which blows up at the DC limit. We will later see that this ratio $\varepsilon_0$ plays a key role in the behaviour of the butterfly velocity. In some sense, such a result is not surprising \cite{Viktor, Dimitrios1}  since in theories with anisotropic scalings, one also observes a violation of the KSS bound \cite{KSSviolationAniso,KSSviolationHighD,Jain:2015txa}. As shown in  \cite{LandsteinerViscosity}, in the model we consider, the viscosity along the anisotropic direction $\eta_\mypar$ violates the KSS bound (see figure \ref{figeta}). It is important to note that the ratio between the $\perp$ quantities and their $\mypar$ relatives is always fixed by the anisotropic parameter defined previously,
\begin{equation}
\frac{\sigma_{\perp}}{\sigma_\mypar} \,=\, \frac{\eta_{\perp}}{ \eta_\mypar}\,=\,1\,+\,\varepsilon_0\,.
\label{theratio}
\end{equation}
We will next see that this will still be true for the butterfly velocities $v_B^2$ and will ultimately be responsible for the violation of the maximization hypothesis. We show the behavior of the anisotropy parameter $\varepsilon_0$ is a function of $\bar{M}$ in figure \ref{figAN}.  As already discussed, the anisotropy parameter is peaked around the quantum critical point and it blows up at $T=0$ following Eq. \eqref{blow}.

\section{Quantum Chaos \& Universality}
\label{sec:chaos}
In this section, we compute the butterfly velocity for the above holographic model. After obtaining a general expression of $v_B$ in terms of the near-horizon data for a given background, we (numerically) solve it near the quantum phase transition. Consider an \textit{anisotropic} black brane metric 
\beq
ds^2 = - f(r) dt^2 + \frac{dr^2}{f(r)} + \sum_{\eta} h_{(\eta)}(r) d\vec{x}_{(\eta)}^{\,2} 
\, .
\eeq
Here $\eta$ (not to be confused with viscosity) counts the number of different warp factors, $h_{(\eta)}(r)$, present in the $\Sigma_{\eta} = \{ \vec{x}_{(\eta)} \}$ sub-manifold of the above background; thus, $D = \sum_\eta d_\eta$, where $d_\eta = \text{dim}(\Sigma_\eta)$. The growth of the commutator in Eq. \eqref{OTOCdef} can be studied in holography by perturbing a black hole with a localized operator $\mathcal{V}(\vec{x}, t)$ \cite{Roberts:2014isa,DanRoberts16}. After a sufficiently long time, ($t > t_r = \beta$) the backreaction of this perturbation grows enormously, giving rise to a shockwave profile, $\psi(\vec{x}, t)$, spreading at a speed $v_B$. Before the perturbation has been completely scrambled ($t < t_s + |\vec{x}|/v_B$), the OTOC behaves as $\sim \psi(\vec{x}, t)^2$. In Appendix \ref{vbAPP} we solve the shock-profile for the above background and obtain the butterfly velocities for an anisotropic AdS background. Note that in an anisotropic background, the velocity of the shockwave-front will depend on the spatial sector $\Sigma_\eta$, and the full profile $\psi(\vec{x}, t)$ can be approximated as a product of the shock-profile of each sector. Doing so, we obtain 
\bea
v_B^{(\zeta)} = \frac{\lambda_L }{\mu \sqrt{h_{(\zeta)}(r_0)} } \quad , \quad \mu^2 =   2 \,\pi\, T \, \sum_{\eta}  \frac{d_\eta}{2} \, \frac{ h'_{(\eta)}(r)}{h_{(\eta)}(r)} \Big\rvert_{r_0} \, .
\label{eq:generalVB}
\eea
Note that $1/\mu$ defines a theory-dependent, dimensionless IR length-scale in the problem, a  screen length over which the shock-profile (exponentially) decays, see Eq. \eqref{eq:ShockProfile}. This quantity plays an important role in our discussion and below we analyze this further.

\begin{figure}[htp]
\centering
\includegraphics[width=0.48 \columnwidth , height= 0.42 \columnwidth]{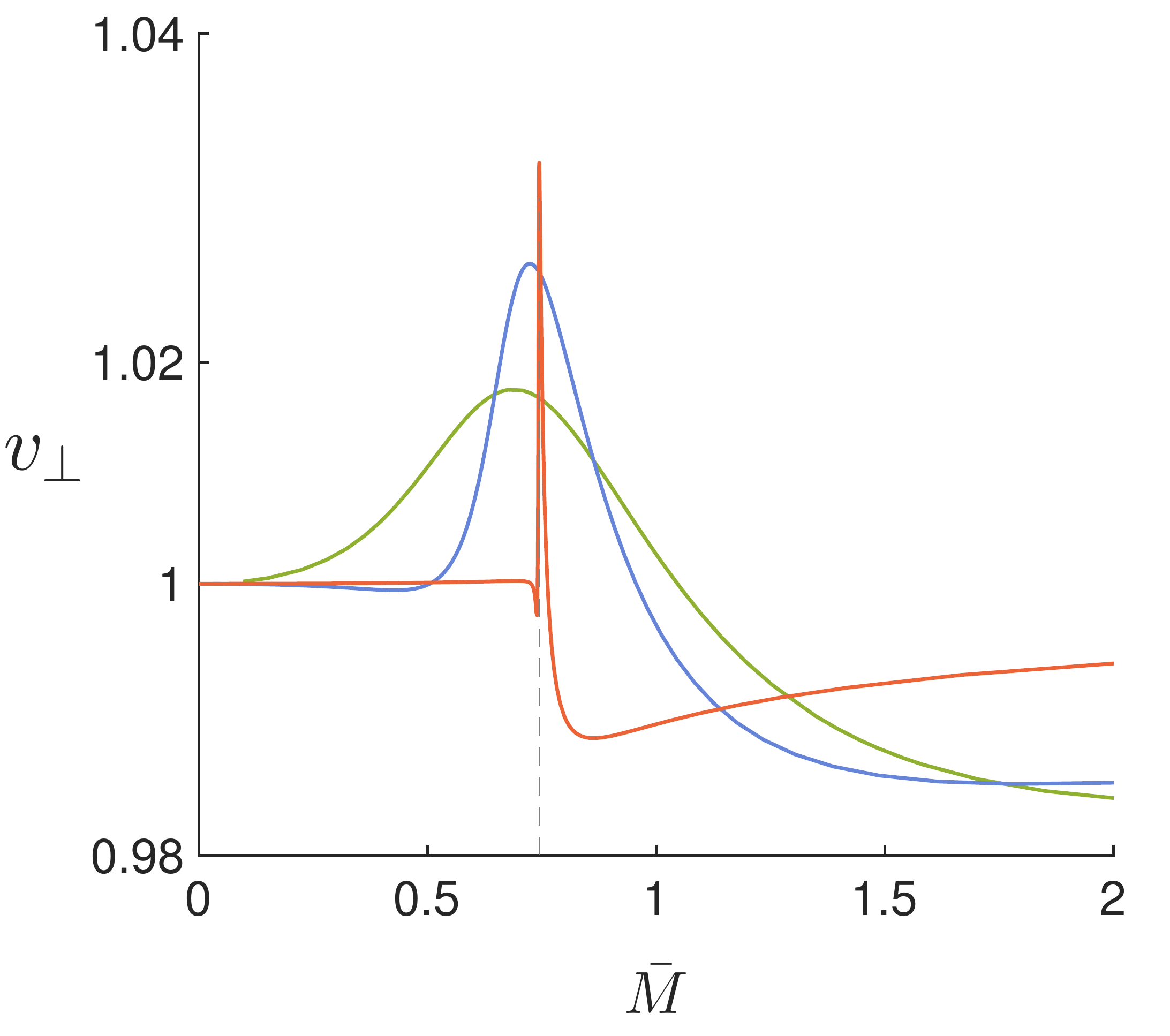}
\includegraphics[width=0.48 \columnwidth , height= 0.42 \columnwidth]{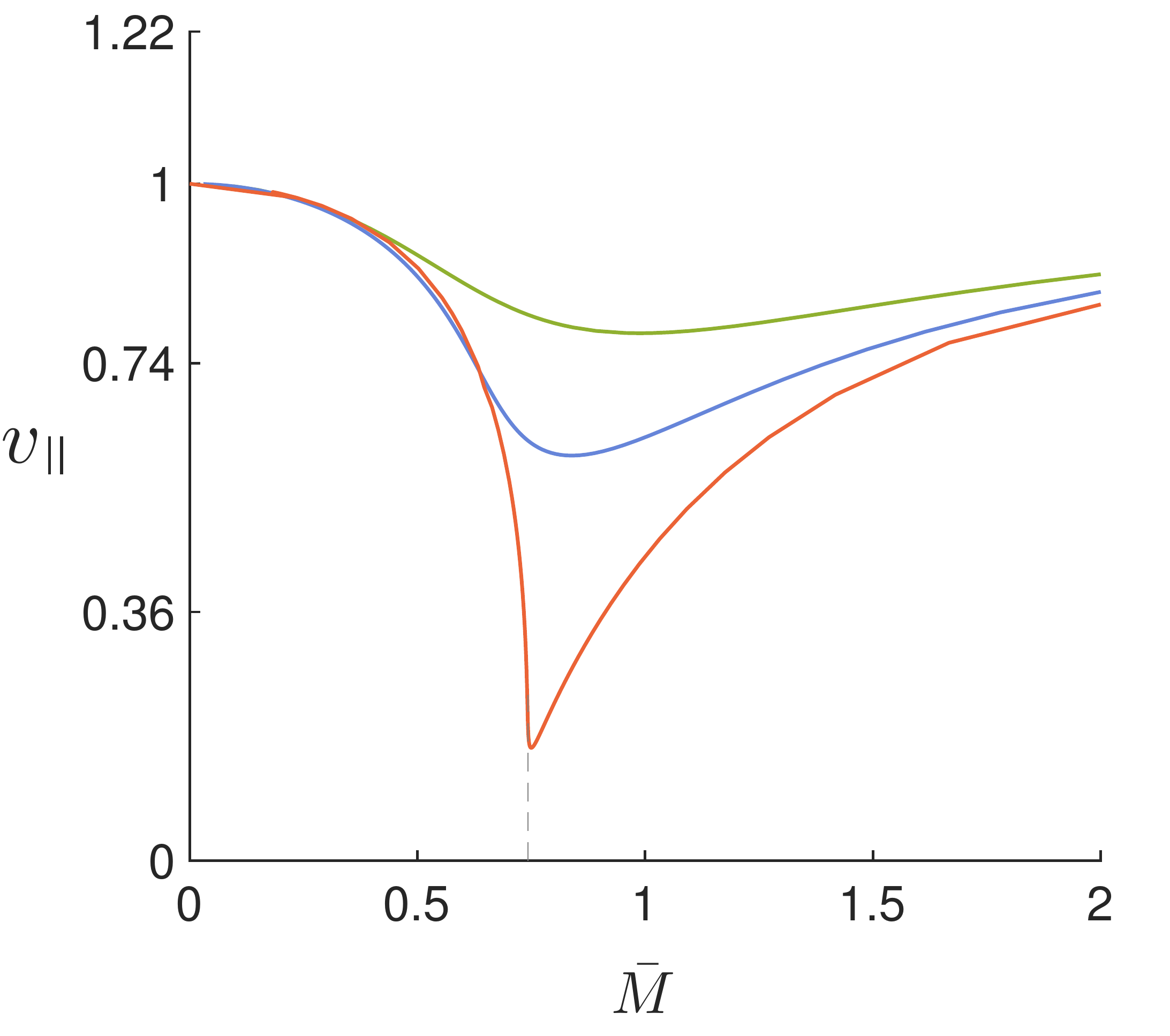}
\caption{Butterfly velocities as function of dimensionless mass $\bar{M}$, for various temperatures $\bar{T}=0.1,0.05,0.005$ (from green to orange). The grid-line is drawn at $\bar{M}_c \simeq 0.74$ in order to show the QCP. As one lowers the temperature the behavior of $v_B$s near the critical point becomes increasingly non-analytic. Note the longitudinal (w.r.t. anisotropy direction) butterfly velocity behaves exactly opposite to its maximization  observed in \cite{ShenQCP}. The $v_B$ values have been normalized  by their asymptotic values at $\bar{M}=0$, that is, $2/\sqrt{6}$. This is obtained from Eq. \eqref{vel}, and is equal to the bound in \cite{MarkMezei}, $v_B^2 = (D+1)/2D$, which is clearly violated\protect\footnotemark\,
 by $v_\perp$ at larger $\bar{M}$.
}
\label{fig:VB}
\end{figure}

An alternative way to express this is through the following near-horizon quantities -- surface gravity, $\kappa = 2 \,\pi\, T$, and the area density of the $r$-slices, which relates the horizon with the entropy density of our dual QFT. We define the density of an $r$-slice which is simply proportional to the area of the spatial surface, $\mathcal{A}^2(r) \sim \Pi_{\eta} \, h^{d_\eta}_{(\eta)} (r) $. Thus $\mu$ is
\beq
{\mu^2} =  \kappa \, \left.  \frac{ \partial}{\partial r} \log \mathcal{A} \,\right \rvert_{r = r_0}\,.
\label{eq:mulog}
\eeq
For the holographic model considered in the previous section, we have one anisotropic direction $z$, that is,  two butterfly velocities. The velocity along the $z$-axis is denoted $v_\mypar$ and that on the $xy$-plane is denoted $v_\perp$. Now we use Eq. \eqref{eq:generalVB} to obtain the butterfly velocities for the background in Eq. \eqref{ansatz}. Since this a holographic theory, the Lyapunov exponent naturally saturates the Maldacena bound \cite{Maldacena:2015waa}, $\lambda_L = 2\pi/ \beta$. In the unit of $\hbar = 1 = k_B$, the maximal Lyapunov exponent is equal to surface gravity, $\lambda_L = \kappa$; however, to avoid ambiguity relating the source of the thermal factor, we continue distinguishing them and write
\begin{gather}
v_{\perp} = \frac{2 \,\pi}{\beta \, \mu \,\sqrt{g_1}}\qquad ,   \qquad v_{\mypar} = \frac{2 \,\pi }{ \beta \,\mu \,\sqrt{h_1}}\, ,\\
{\mu^2}  =  \kappa \left( \frac{g_2}{g_1} + \frac{h_2}{2h_1}  \right)
 = 6 - \left(m^2 + \frac{q^2 A_{z1}^2}{h_1}  \right) \frac{\phi_1^2}{2}  - \lambda \frac{\phi^4_1}{4}  
 \, .
 \label{vel} 
\end{gather}
Here we have used the near-horizon expansion of the metric functions, $g(r) = g_1 + g_2 (r-r_0)$ and $h(r) = h_1 + h_2(r-r_0)$ discussed in Appendix \ref{app1}, which involves $A_z(r_0) \equiv A_{z1}$ and $r_0 \phi(r_0) \equiv \phi_1$. Also, we have set the horizon radius to $r_0 = 1$. As discussed in the previous section, the boundary theory is described by two dimensionless parameters, $(\bar{M}, \bar{T})$. In turn, this fixes two near-horizon quantities, $(\phi_1, A_{z1})$. All other IR variables are functions of $(\bar{M}, \bar{T})$, through $(\phi_1, A_{z1})$. In figure \ref{fig:VB} we numerically obtain the behavior of the butterfly velocities. Although, as noted in \cite{LingQCP}, there is a characteristic behavior of $v_B$s near the critical point; however, there is a clear departure from the result of \cite{ShenQCP} since the velocity along the anisotropic direction seems to attain a local minimum around the critical point, instead of a local maximum. 
\footnotetext{We thank Viktor Jahnke for pointing this out. This bound was observed to be violated in \cite{Viktor, Dimitrios1}.}
The apparent inability of $v_{\mypar}$ to attain a maximum can be traced back to the anisotropic scaling. As before, this can be seen from the ratio,
\beq
\frac{v^2_{\perp}}{v^2_{\mypar}} = \frac{h(r_0)}{g(r_0)}\,=\,1\,+\,\varepsilon_0 \, .
\eeq
Since we observe finite $v^2_{\perp}$ at $g=g_c$, the divergence of this ratio at the critical point causes $v_{\mypar}$ to vanish. In other words, it is the length scale appearing in the formula of the butterfly velocity that sources the deviation from the maximization behavior. Hence, modulo this length scale, $v_B^{(\eta)}$ maximizes only when $\mu$ is minimized. Hence, if we consider the dimensionless information screening length $\mathrm{L}\equiv 1/\mu$ instead, perhaps a universal statement can be made irrespective of the anisotropic scaling of the QPT. In this regard, we conjecture that $\mathrm{L}$, and not the butterfly velocity $v_B$, \textit{maximizes} across a quantum phase transition. Notice that in the isotropic case, the two statements are perfectly equivalent, and therefore the previously conjectured bound holds. Before discussing this more generally, we analyze the asymptotic limits of $\mu^2$ in our system, using Eq. \eqref{vel} as a guide. 

\begin{figure}
\center
\includegraphics[width=0.82 \columnwidth , height= 0.67 \columnwidth]{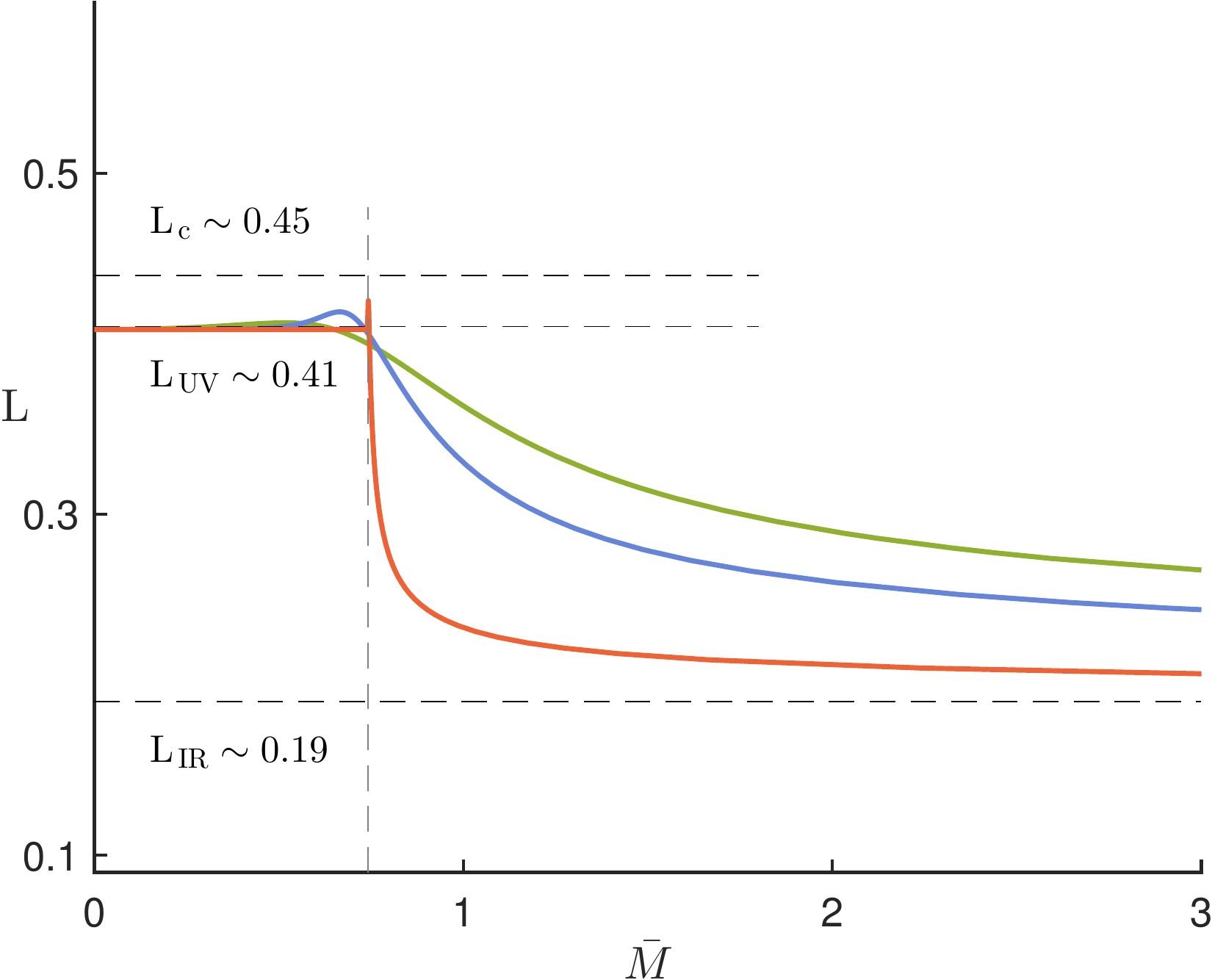}
\rlap{\hspace{-2.15 in}\raisebox{2.4 in}{%
\includegraphics[width= 2.1 in]{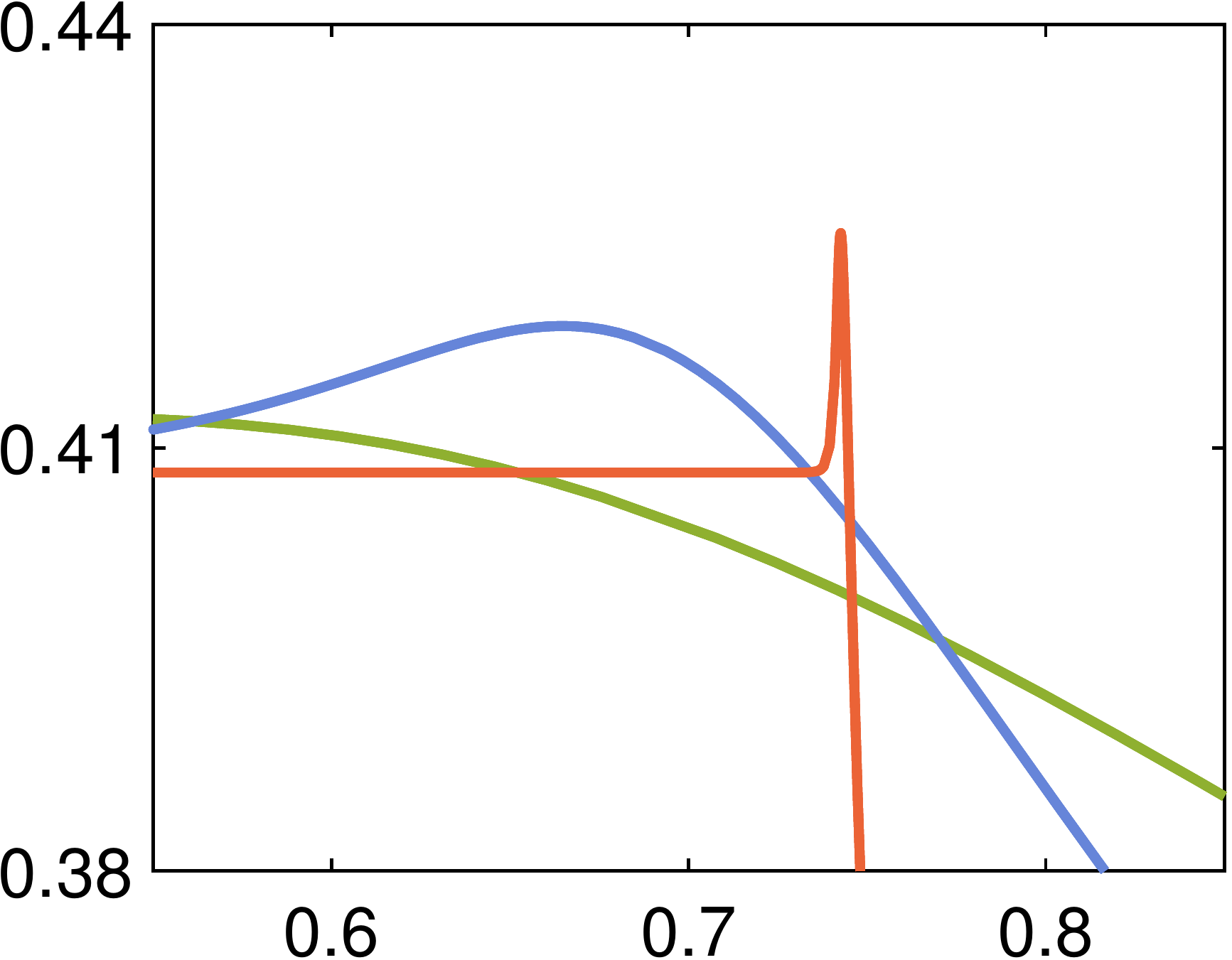}}
}
\caption{Dimensionless scale $\mathrm{L}$ across the quantum phase transition for various temperatures $\bar{T}=0.1,0.05,0.005$ (from green to orange). In the IR limit it asymptotes to $\mathrm{L}_{\, \text{IR}}$ and in the UV this is $\mathrm{L}_{\,\text{UV}}$. The inset zooms into the behavior around the critical point. For low temperature $\mathrm{L}$ maximizes around the critical temperature and reaches the maximum, $\mathrm{L}_{\, \text{c}} = (4+2\beta_0)^{-1/2}$. Since NEC ensures $\beta_0 < 1$, thus $\mathrm{L}_{\, \text{c}}$ is always larger than $\mathrm{L}_{\, \text{IR}}$. In the text we argue for this maximum to be a universal property.}
\label{piccon}
\end{figure}
 
 Firstly, at $\bar{M} = 0$, since there is no perturbation, we have $\mymu{UV} = \sqrt{6} \simeq 2.45$. The factor of $6$ is simply twice the spatial-dimension of the boundary CFT, $2D = 2\sum_{\eta} d_\eta$, which also fixes the butterfly velocity of a $d$-dimensional Schwarzschild black hole background \cite{StringyScramble}.  At zero temperature, as $\bar{M}$ is increased, until one crosses $\bar{M}_c$, there is no condensate, causing $\mu^2$ to stay unchanged. At the critical point (using Eq. \eqref{bkg:Lifs} for the critical background) we have $\mymu{c} = (4 + 2 \beta_0)^{1/2} \simeq 2.19$. As discussed before, NEC forces $\beta_0 < 1$.  In turn this causes $\mu^2$ to sharply decrease at the critical point. For an isotropic system ($\beta_0=1$), one observes no transition in $\mu^2$. This sharp transition at the critical point for $\beta_0 \neq 1$ smears out becoming a cross-over behavior at finite temperature.  A final question is whether or not $\mu^2$ monotonically decreases after the transition or if it increases. The IR asymptotic value of $\mu^2$, using the data of Eq. \eqref{bkg:Ins}, is $\mymu{IR} = (6 + 9/4 \lambda)^{1/2} \simeq 5.34$. Clearly this is larger than $\mymu{c}$; in fact it is bound to be larger than $\mymu{UV}$ as well since $\lambda$ is always positive. At finite temperature this asymptotic value softens but stays larger than the critical value for low  enough temperature. We plot the behavior of $\mu$ in figure \ref{piccon} which conforms to our inference and conforms to
\beq
\mymu{c} \leq \mymu{UV} \leq \mymu{IR} \quad \text{or,} \quad \mathrm{L}_{\,\text{c}} \geq \mathrm{L}_{\,\text{UV}} \geq \mathrm{L}_{\,\text{IR}}   \, .
\eeq

Now, in the spirit of \cite{TopDownWeyl}, we attempt to understand whether this conclusion remains valid if the boundary operator assumes any other scaling dimension. This discussion is confined  just to the insulating phase since the scalar deformation operator condenses only for large $\bar{M}$. In other words, when the second- or higher- order terms in $\mu^2$ are turned on in Eq. \eqref{vel}. We focus on the behavior of $\mu$ at low temperature, and when $\bar{M} - \bar{M}_c \ll 1$, so that we can simplify our treatment by using the scalar hair $\phi_1$ as a perturbation parameter.  Also, since away from the critical point, $\mu$ behaves analytically and monotonically so as to establish our lower-bound conjecture, it suffices to justify that $\mu$ starts increasing as one enters slightly into the insulating phase. The coefficient of $\mathcal{O}({\phi_1^2})$ term is simply the effective mass of the scalar hair, $m^2_\text{eff} = m^2 + \textsl{g}^{zz} q^2 A_z^2$. Since at low temperature $\textsl{g}^{zz} = 1/h_1  \rightarrow 0$ at the QPT, we first consider $m^2$ only. At this order, $\mu^2 \approx 6 - m^2 \phi_1^2/2$, and only for $m^2 <0$ one has increasing $\mu$. Recall \cite{KlebWitten} that the mass of a bulk scalar field is fixed by the scaling dimension of the dual boundary operator as $m^2 = \Delta(\Delta - d)$. The BF instability prevents this mass from becoming smaller than $m_{\text{BF}} = -d^2/3$ (in this case, $m_{\text{BF}} = -4$). For our conjecture, $m^2 < 0$ is true as long as $\Delta < d$, or the perturbation is relevant. It should be noted that this is a fundamental requirement in order to generate a QPT, since by perturbing a UV with an irrelevant operator, one can never generate a non-trivial RG flow towards an IR fixed point. This is indeed the case as noted in the numerical studies of \cite{TopDownWeyl}. Thus, irrespective of the scaling dimension of the boundary deformation operator, one can define a lower bound on the length scale of information scrambling, which is fixed by the CFT$_d$. For a non-relativistic CFT$_d$ with a scaling anisotropy $\beta_0$, along a $D_\mypar$-dimensional sub-space ($D_\mypar=D - D_\perp $), the upper bound is (using Eq.~\eqref{eq:mulog} for a generic background)
\beq
2 \mathrm{L}\leq \frac{1}{D\,+\,(\beta_0-1)\,D_\mypar }  \equiv 2 \mathrm{L}_{c} \,,
\label{eq:bound}
\eeq
and the equality is saturated exactly at the quantum critical point\footnote{
Since the anisotropic geometry turns out to be the critical geometry in the above model, the saturation happens at the QCP leading to the violation of the maximization-result. However, a system exhibiting such geometries in the UV or IR might saturate this bound away from the QCP. Thus, the significance of the bound should not necessarily be attached to quantum criticality but rather should be seen more as a universal feature of the near-horizon IR geometry. We thank Elias Kiritsis for discussing this issue.
}, $g=g_c$ as illustrated in the figure above. Note that ultimately it is the NEC that restricts $\beta_0$ to be less than one, and hence, makes the critical value $\mathrm{L}_{\text{c}}$ larger as compared to any other asymptotic value. In the case of isotropy, the maximum on the information screening length $\mathrm{L}$ becomes translated to the maximum of the butterfly velocity $v_B$ since $v_B\,\sim\,\lambda_L\,\mathrm{L}$. Nevertheless, as we showed, in the presence of anisotropy ($\beta_0 \neq 1$), the statement about the butterfly velocity does not hold anymore and it has to be replaced by the behavior of the dimensionless information screening length $\mathrm{L}$.

\section{Conclusion}
\label{disc}
Throughout this work, we studied the onset of quantum chaos on an anisotropic quantum phase transition in a holographic bottom-up model. In particular, we focused on the behavior of the butterfly velocities in the quantum critical region and across the quantum phase transition. We observed a disagreement with the results proposed in \cite{ShenQCP}. More precisely, the butterfly velocity along the anisotropic direction does not develop a maximum but rather a minimum at the quantum critical point. We reiterate the similarity of our conclusions with the violation of the Kovtun--Son--Starinets (KSS) lower bound on the viscosity to entropy density ratio \cite{KSSviolationAniso,Jain:2015txa}. In either cases, the presence of the anisotropic scaling, $\beta_0$ seems to play an identical role. The viscosities have indeed been computed \cite{LandsteinerViscosity} within the holographic model we considered and, as expected and already mentioned, the $\eta/s$ ratio along the anisotropic direction violates the KSS bound, recall figure \ref{figeta}.

As a remedy, we propose an improved conjecture which also holds in the presence of anisotropy, and is stated in Eq.~\eqref{eq:bound}. This involves a length scale, $\mathrm{L}$, from the bulk perspective which can be computed using Eq.~\eqref{eq:mulog}. For the boundary theory this may be indirectly extracted by measuring the ballistic growth of a local perturbation through the OTOC and combining this with the measurement of various transport properties such as viscosity or conductivity along specific anisotropic directions. This is needed since the factors $g(r_0)$ or $h(r_0)$ can only be made relevant to the boundary theory through these quantities, such as in  Eq.~\eqref{eq:CondForm}. In an anisotropic case, we observe $\mathrm{L}_{\,\text{c}} \geq \mathrm{L}_{\,\text{UV}} \geq \mathrm{L}_{\,\text{IR}} $; however for 
the isotropic case we do not expect $\mathrm{L}$ to have a local maximum at the critical point, that is $\mathrm{L}_{\,\text{c}} = \mathrm{L}_{\,\text{UV}}$. It would be interesting to understand the physics behind this $\mathrm{L}$ more precisely, especially to see if the emergence of this length scale in a strongly correlated theory can be better understood without making any reference to AdS/CFT.

\section*{Acknowledgments}
We thank Panagiotis Betzios, Alessio Celi, Thomas Faulkner, Karl Landsteiner, Yan Liu, Napat Poovuttikul, Valentina Giangreco Puletti, for useful discussions and comments about this work. We thank Ben Craps, Dimitrios Giataganas, Viktor Jahnke and Elias Kiritsis for valuable and constructive comments on the first version of this paper.  We are grateful to Wei-Jia Li for reading a preliminary version of the draft. We acknowledge support from Center for Emergent Superconductivity, a DOE Energy Frontier Research Center, Grant No. DE-AC0298CH1088.  We also thank the NSF DMR-1461952 for partial funding of this project. MB is supported in part by the Advanced ERC grant SM-grav, No 669288. MB would like to thank Marianna Siouti for the unconditional support. MB would like to thank University of Iceland for the ''warm'' hospitality during the completion of this work and Enartia Headquarters for the stimulating and creative environment that accompanied the writing of this manuscript.

\appendix
\section{The Holographic Background}\label{app1}
\begin{figure}[htbp]
 \center
 \includegraphics[width=7.4cm]{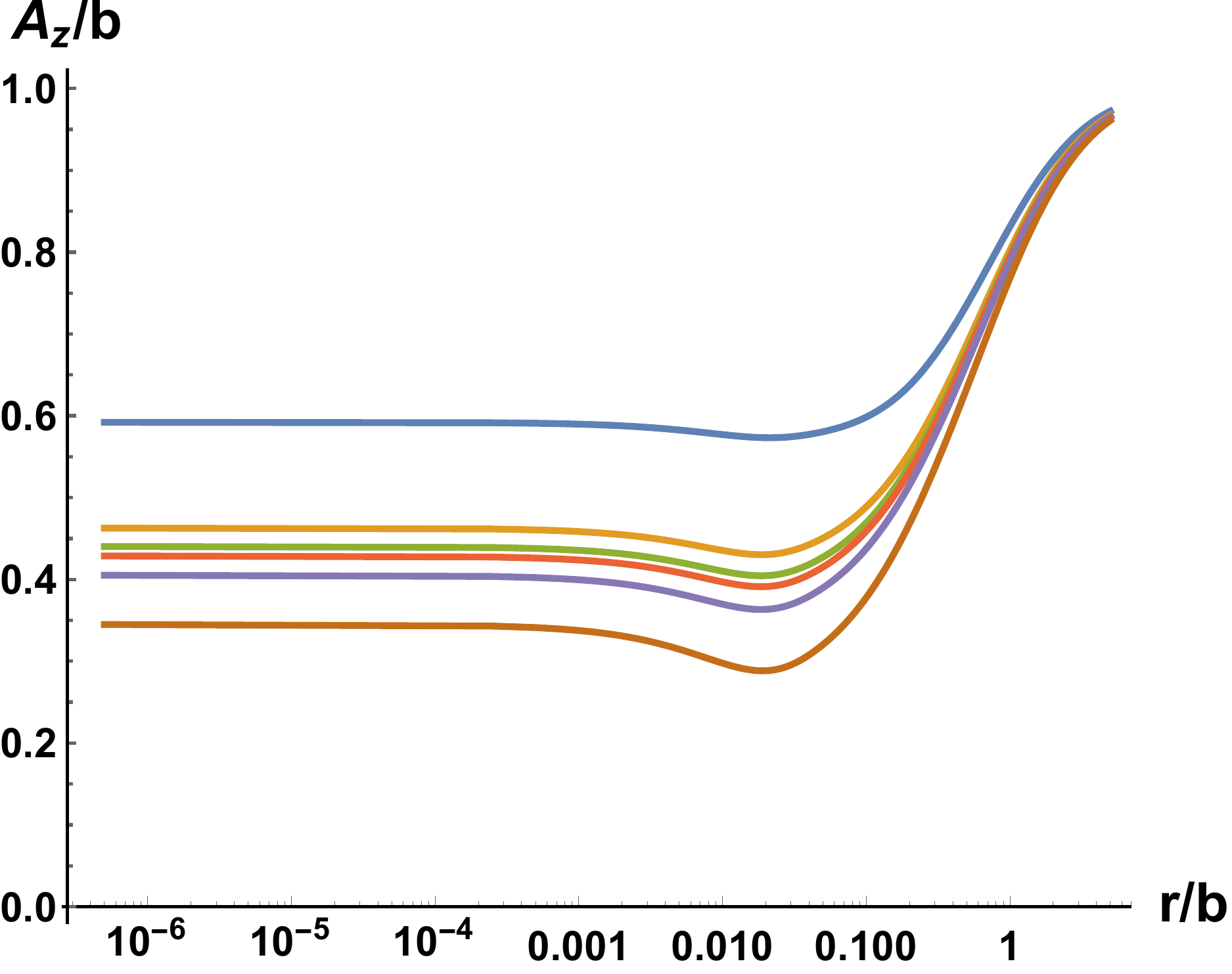}
 \quad
 \includegraphics[width=7.4cm]{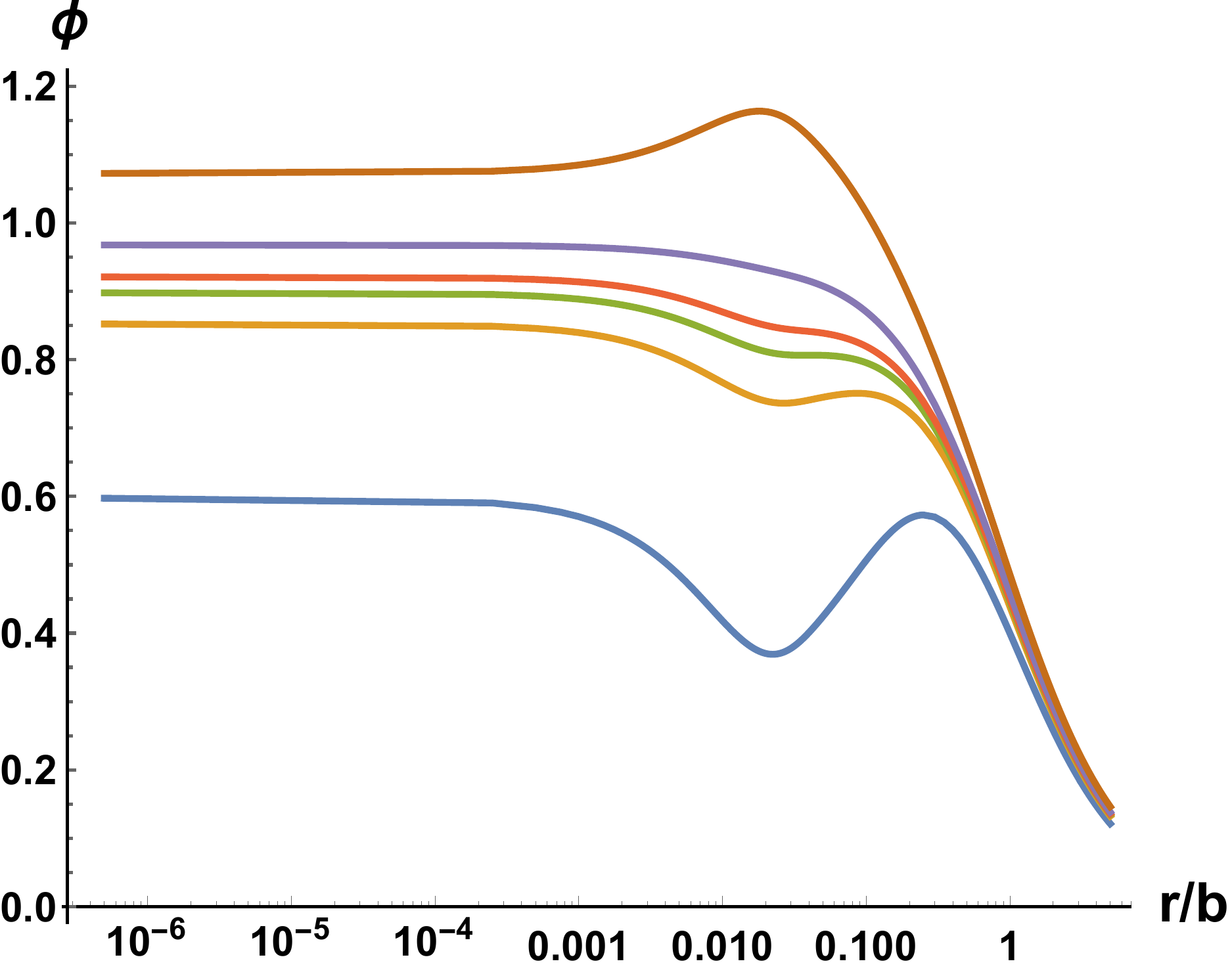}
\caption{Log-linear plot of the bulk profiles for the gauge field $A_z(r)$ and the scalar field $\phi(r)$ at $\bar{T}=0.05$. The various colors (from blue to brown) are $\bar{M}=0.66,0.724,0.736,0.743,0.757,0.8$. The phase transition can be seen from the a large shift og the near-horizon values of the bulk fields when $\bar{M}$ exceeds $0.744$.}
 \label{figbulk}
 \end{figure}

\begin{figure}
 \center
 \includegraphics[width=4.8cm]{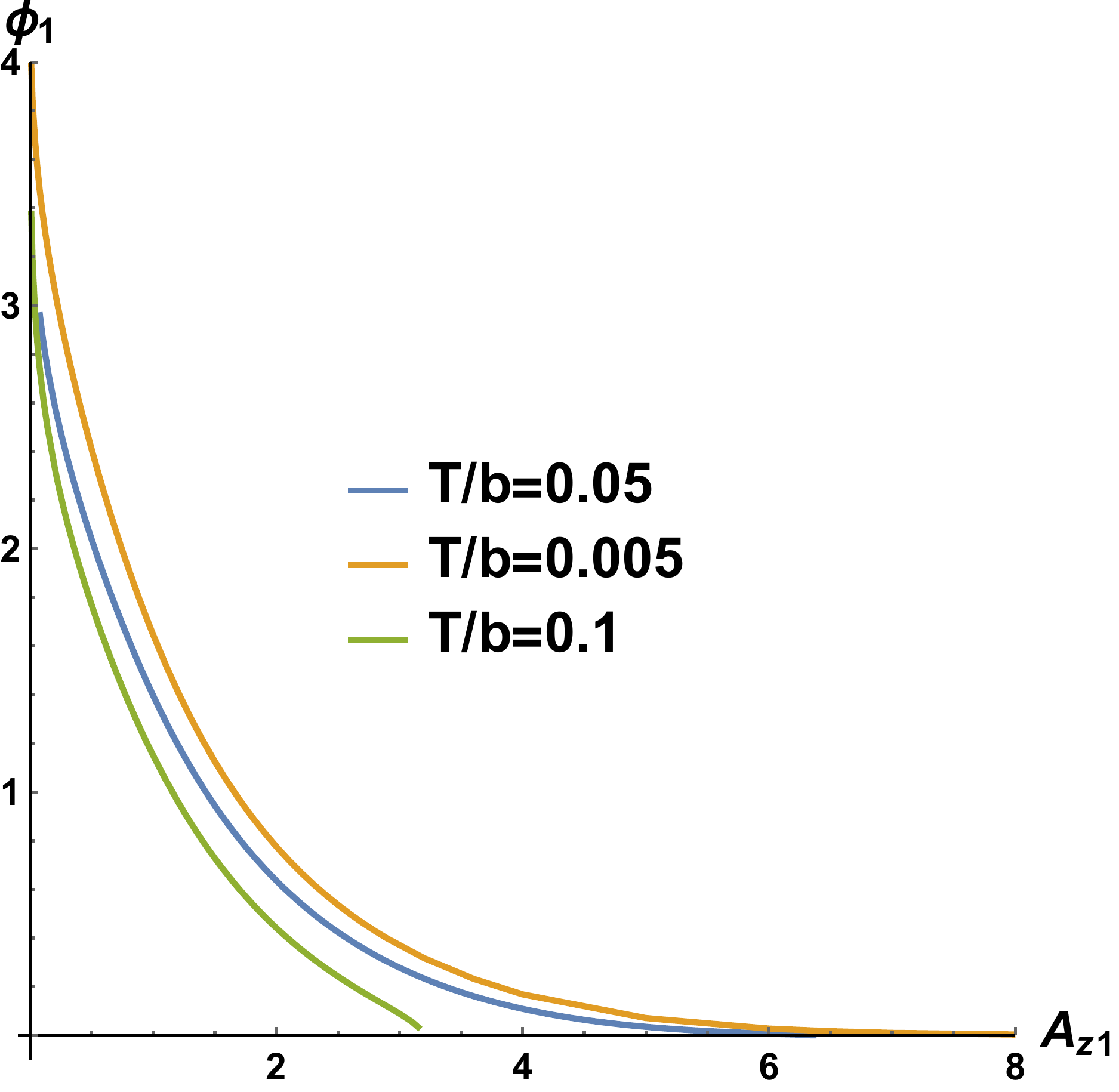}
 \quad
 \includegraphics[width=4.8cm]{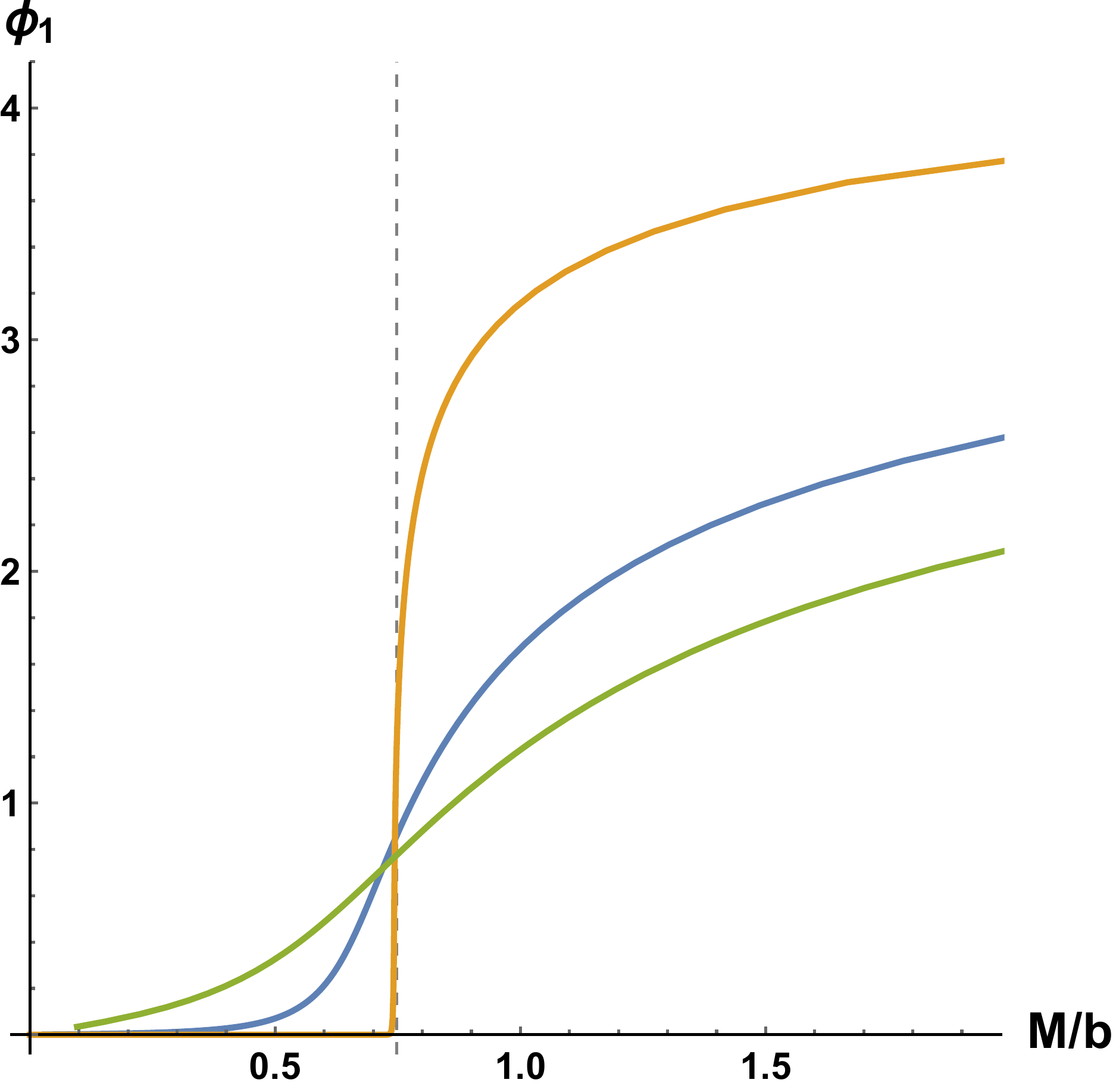}
\quad
\includegraphics[width=4.8cm]{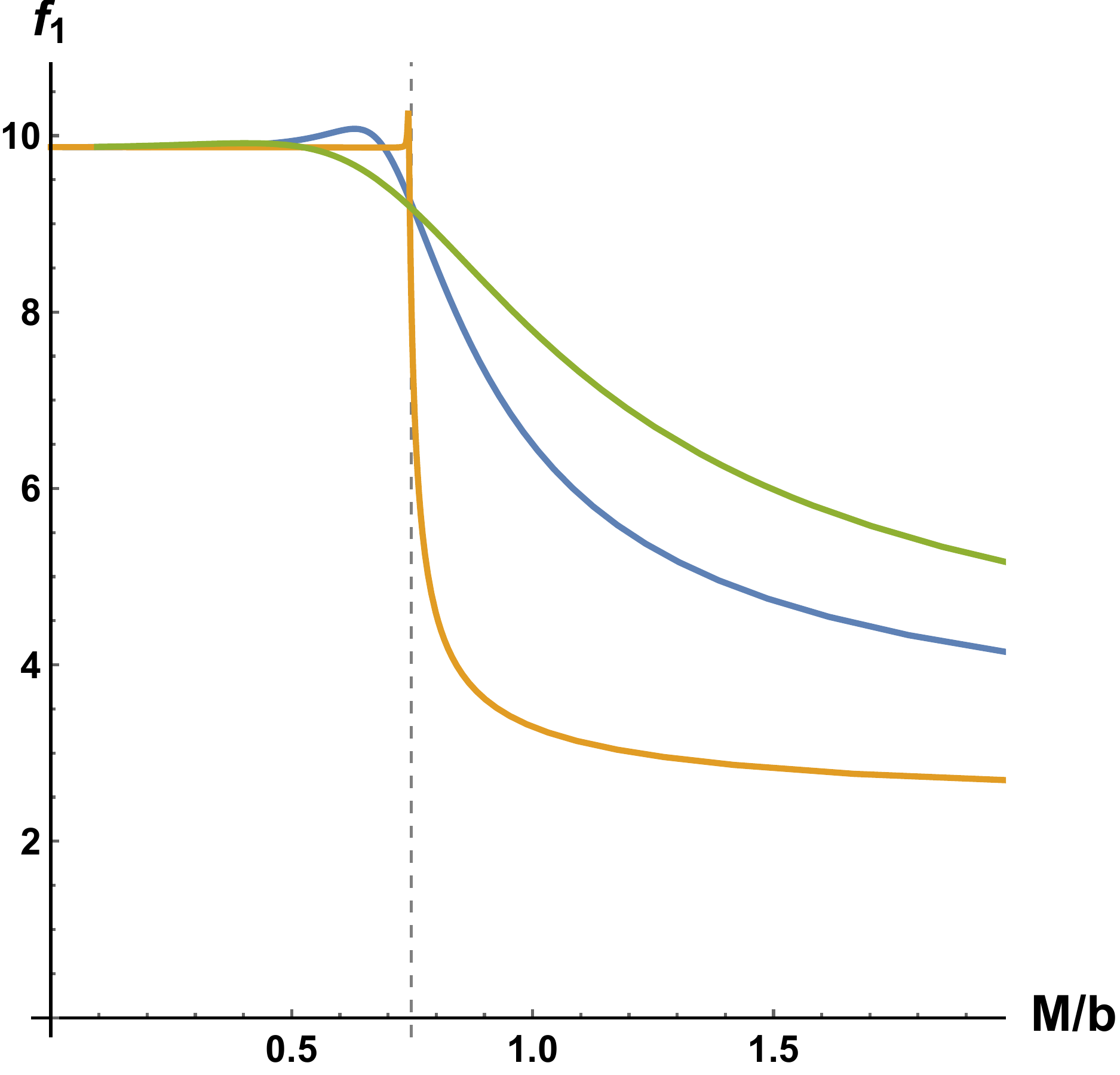}
\caption{Numerical details for the results presented in the main text for $\bar{T}=0.1,0.05,0.005$. {Left:} the values of $({A_z}_1,\phi_1)$ for the horizon shooting. {Center: } the value of $\phi_1$ in function of $\bar{M}$. {Right:} the value of $f_1$ in function of $\bar{M}$.}
 \label{fignum}
 \end{figure}

We discuss some more details about the gravitational background here and some aspects of the pertaining numerics. We follow closely \cite{Landsteiner:2015pdh}. The equations of motions derived combining the action in Eq. \eqref{action} with our ansatz in Eq. \eqref{ansatz} are (note in order to be consistent with the notations in Landsteiner \textit{et al.} we have switched $f \rightarrow u, \, g \rightarrow f$):
\begin{subequations}
\begin{align}
&u'' +\frac{h'}{2 \,h}\, u'\, - u\left(\frac{f''}{f}+\frac{f'\, h'}{2 \,f\,
   h}\right)=\,0\,, \\
&\frac{f''}{f}+\frac{f'\, u'}{f\,u}-\frac{{f'}^2}{4
   f^2}+\frac{u''}{2
   u}-\frac{{A_z'}^2}{4 \,h} + \frac{m^2 \phi^2}{2 u}-\frac{q^2 A_z^2 \phi^2}{2 h
   u}+\frac{\lambda  \phi^4}{4 u}-\frac{6}{u}+\frac{1}{2} {\phi
   '}^2\,=\,0\,,  \\
& \frac{{A_z'}^2}{4 h} - \left(\frac{f'}{2 f\,u}+\frac{h'}{4 \,h\,
   u}\right)u'-\frac{f'\, h'}{2 f\, h}-\frac{{f'}^2}{4
   f^2}- \left({m^2}+\frac{q^2 A_z^2 }{h}+\frac{\lambda  \phi^2}{2} \right) \frac{\phi^2}{2u}+\frac{6}{u}+\frac{1}{2} {\phi '}^2=0 \, , \\
   &A_z''+A_z'(r) \left(\frac{f'}{f}-\frac{h'}{2
   h}+\frac{u'}{u}\right)-\frac{2 q^2 A_z\,\phi^2}{u}=0\,,\\
& \phi ''+\phi '
   \left(\frac{f'}{f}+\frac{h'}{2
   h}+\frac{u'}{u}\right)-\frac{\lambda  \phi^3}{u} + \left(-\frac{q^2\,A_z^2}{h\,u}-\frac{m^2}{u}\right) \phi =0\,.
\end{align} \label{backEOMs}
\end{subequations}
Here the primes denote derivative with respect to the radial-coordinate. We want to nnumerically integrate the system of equations \eqref{backEOMs} from the horizon $r= r_0$ to the boundary $r=\infty$. In order to do so we first try to find the asymptotic behavior of the solutions near the IR boundary (horizon) and UV (conformal) boundary. Close to the UV boundary, the bulk fields have the following leading order asymptotic expansion:
\begin{align}
&u\,=\,r^2\,+\,\dots\,,\quad f\,=\,r^2\,+\,\dots\,,\quad h\,=\,r^2\,+\,\dots\,,\quad A_z\,=\,b\,+\,\dots\,,\quad \phi\,=\,\frac{M}{r}\,+\,\dots\,.
\end{align}
Note that we have rescaled the boundary values of the three different metric functions to unity, such that the boundary field theory depends only on the following free parameters, $T, b, M$. The removal of the boundary values of the metric is achieved by invoking the following (three) scaling symmetries 
\begin{enumerate}
\item $(x,y)\rightarrow a(x,y),\,\,f\rightarrow a^{-2}f$ ;
\item $z\rightarrow a z,\,\,h \rightarrow a^{-2}h,\,\,A_z\rightarrow a^{-1}A_z$ ;
\item $r\rightarrow a r,\,\,(t,x,y,z)\rightarrow (t,x,y,z)/a,\,\,(u,f,h)\rightarrow a^2(u,f,h),\,\,A_z \rightarrow a A_z$ .
\end{enumerate}
Owing to there symmetries we only have two dimensionless scales, $\bar{T}$ and $\bar{M}$, which control the entire of the solution space. The near-horizon expansion up to $\mathcal{O}(r-r_0)$ can be written as 
\begin{align}
\begin{split}
&u \simeq 4\,\pi\, T\,(r-r_0)+u_2\,(r-r_0) \,\,,\,\, f\simeq f_1+f_2\,(r-r_0)\,\,,\,\,\, h\simeq h_1+h_2\,(r-r_0)\,\,, \\
&A_z\simeq {A_z}_1+{A_z}_2\,(r-r_0) \,\,,\,\, r \phi \simeq \phi_1\,+\,\phi_2\,(r-r_0) \, .
\end{split}
\end{align}
Here $Az_1$ and $\phi_1$ are the only free parameters, being controlled by the boundary data $\bar{T}$ and $\bar{M}$. From now onward, we also set the horizon radius to $r_0=1$. In summary, while the horizon data are $(T,r_0,f_1,h_1,{A_z}_1,\phi_1)$, using the (three) scaling symmetries they get reduced to $(T,{A_z}_1,\phi_1)$. At the conformal boundary they take the form of $(T, M, b)$. We can now use shooting to construct the numerical background on the 2D plane of $(\bar{M},\bar{T})$. An example of the bulk profiles for the $A_z(r)$ and $\phi(r)$ fields is shown in figure \ref{figbulk} and figure \ref{fignum}.

\section{Butterfly Velocities in Anisotropic Backgrounds}
\label{vbAPP}
Here we set up the shock wave equation in a generic anisotropic (in the spatial field theory directions) background with constant curvature. For this we closely follow the derivations presented in \cite{BlakePRD16, Sfetsos:1994xa, DanRoberts16}. Consider the following $d$-dimensional background with a black hole 
\beq
ds^2_{(0)} = - a(r) f(r) dt^2 + \frac{dr^2}{b(r)f(r)} + \sum_{\eta} h_{(\eta)}(r) d\vec{x}_{(\eta)}^{\,2} 
\, .
\label{eq:GenericBkg}
\eeq
Here $\eta$ counts the number of different warp factors, $h_{(\eta)}(r)$, present in the $\Sigma_{\eta} = \{ \vec{x}_{(\eta)} \}$ sub-manifold of the above background. The treatment of Sfetsos confines to $\eta=1$, however, here we are interested in the case when $\eta > 1$. The black hole (or black brane) horizon is assumed to be located $r_0$, such that $f(r_0)=0$ with non-vanishing $a(r_0)$ and $b(r_0)$. The temperature of the black hole is, $4 \pi T =  2 \kappa= f'(r_0) \sqrt{a(r_0) \, b(r_0)}$, here $\kappa$ is the surface gravity. The background is assumed to be sourced by a stress tensor, $T_{\mu \nu}^{(0)}$. For further simplifications we first move to tortoise coordinate,
\begin{gather}
ds_{(0)}^2 = a(r) f(r) \left(dr_*^2 - dt^2\right) + \cdots \, ,  \\
r_*(r) = \int_{r_0}^r \frac{dr'}{f(r') \sqrt{a(r')\,b(r')}} \approx \frac{1}{4 \pi T} \ln \frac{r- r_0}{r_0}  \, .
\end{gather}
In the last line we've done a near-horizon expansion of $r_*$ which is justified since $r_*(r_0)$ blows up. Next we move to Kruskal coordinate by exponentiating the null coordinates of $t-r_*$ space,
\bea
u = e^{2 \pi T \left( r_* - t \right) } \quad , \quad v =  e^{2 \pi T \left( r_* + t \right) }  \implies  r_*  = \frac{1}{4 \pi T }\, \ln(uv)  \quad , \quad t = \frac{1}{4 \pi T }\, \ln \frac vu
\eea
In this coordinate the horizon is at $uv = 0$ and the boundary is at $uv=-1$. The black hole singularity is at $uv =1$. The above relation can be used to express the background in Kruskal coordinates
\bea
ds_{(0)}^2 = 2 A(uv) du dv + \sum_{\eta} h_{(\eta)}(u v) d\vec{x}_{(\eta)}^{\,2}  \quad , \quad 
2A(uv) =  \frac{a(r) f(r) }{(2 \pi T)^2} e^{ - 4\pi T r_*} \, .
\label{eq:KruskalMetric}
\eea
We will need the following relations later, $h'(0) = r_0 h'(r_0)$, and using near-horizon expansion of $f(r)$ we have, $2A(0) = \frac{r_0}{(2 \pi T)^2} {a(r_0)f'(r_0)}$ and  $2A'(0) = \frac{r_0^2}{(2 \pi T)^2} \left( a(r) f'(r) \right)' \rvert_{r_0}$. One can think of the above background is being generated from stress tensor $T^{(0)}$ by using Einstein equation, $G^{(0)}_{\mu \nu} = 8 \pi T^{(0)}_{\mu \nu}$, where $G_{{\mu \nu}}^{(0)}$ is the Einstein tensor corresponding to $ds^2_{(0)}$ and
\bea
T^{(0)} = T^{(0)}_{uv} \, dudv + T^{(0)}_{uu} \, du^2 + T^{(0)}_{vv} \, dv^2 + T^{(0)}_{\eta \eta} \, d\vec{x}_{(\eta)}^{\,2}  + T^{(0)}_{u \eta}\, du dx^\eta \, .
\eea
Starting from Eq. \eqref{eq:KruskalMetric} we now obtain the butterfly velocity. For that we perturb our background with a point particle that is released from $\vec{x}=0$ at time $t_w$ in the past. The particle is localized onn the $u=0$ horizon but moves in the direction of $v$ with light speed. For late time, $t_w > \beta$ its energy density can be written as \cite{Drayt'Hooft85}
\bea
T^{p}_{uu} = E_0 e^{\frac{2 \pi}{\beta} t_w} \delta(u) \delta(\vec{x})
\eea
We want to compute the backreaction of this stress tensor on our background. This can be done perturbatively for a small energy density. One can start with an ansatz solution that $v$ gets shifted by $\psi(\vec{x})$ only for $u>0$, $v \rightarrow v + \Theta(u) \psi(\vec{x})$. This new geometry is the shockwave geometry and we want to solve for $\psi(\vec{x})$, that is the shockwave. By relabeling $v$, we replace $dv \rightarrow dv - \delta(u) \psi(\vec{x}) du$. Plugging this in the above metric we obtain the perturbed metric
\bea
ds^2_{(1)} = - 2 A(uv)  \delta(u) \psi(\vec{x}) du^2 \, ,
\eea
and the stress tensor is (along with $T^p$)
\bea
T^{(1)} =\left( T^p_{uu} - T_{uv}^{(0)} \delta(u) \psi(\vec{x}) \right) \, du^2 - 2 T_{vv}^{(0)} \delta(u) \psi(\vec{x})\, dudv
\eea
Since $ds_{(1)}^2$ doesn't generate finite Einstein tensor, $G^{(1)}_{uv} = 0$, we can demand $\delta(u) T_{vv}^{(0)} = 0 = \delta(u) G_{vv}^{(0)} $. There remains only one relevant Einstein equation that gives rise to the shock wave equation (which is subject to the previous contstraint)
\begin{gather}
G_{uu}^{(1)} = 8 \pi T^p_{uu} - \delta(u) \psi(\vec{x})  G_{uv}^{(0)} \, . 
\\ \text{Or,} \quad
\sum_{\eta} \left( \frac{A(0)}{h_{(\eta)}(0)} \Delta_{(\eta)} - \text{dim}(\eta) \frac{h'_{(\eta)}(0)}{2h_{(\eta)}(0)} \right) \psi(\vec{x}) =  8 \pi E_0 e^{\frac{2 \pi}{\beta} t_w} \delta(\vec{x})\, ,
\\ \implies 
\left(  \Delta_{(\zeta)} - M^2_{\zeta} \right) \psi(x_i^{(\zeta)}) =  \frac{16 \pi E_0 h_{(\zeta)}(0)}{2A(0)} e^{\frac{2 \pi}{\beta} t_w} \delta(x_i^{(\zeta)})\, , 
\\ \text{where,} \quad M_{\zeta}^2 =  {h_{(\zeta)}(0)} \sum_{\eta}  \text{dim}(\Sigma_\eta) \frac{h'_{(\eta)}(0)}{2A(0) h_{(\eta)}(0)}  \label{eq:KruskalMass} 
\end{gather}
In the second last line, assuming linear order, we have divided the solution space into different anisotropy sectors, labeled by $\zeta$. Clearly, for the isotropic case, $\eta=1=\zeta$, one recovers the shock equations of \cite{BlakePRD16, DanRoberts16}, with dim$(\Sigma_\eta) = d-2$. Also if the field theory living at a constant $r,t$-slice is curved then the shock front is no longer planar but depends on the curvature of the spatial slice, thus its dynamics involves curved space Laplacian, $\Delta_{(\eta)}  \equiv \frac{1}{\sqrt{g^{(\eta)}}} \, \partial \left( \sqrt{g^{(\eta)}} \, g^{(\eta)} \, \partial \right) $, rather than the flat space Laplacian used above. This affects the spatial-profile of the shock but not its speed, that is the butterfly velocity \cite{huangVb17}. We want to solve this equation, which is equivalent to solving the Green's function of the flat space Laplacian. At very long distance ($x \gg M_{\zeta}^{-1}$) the solution becomes
\bea
\psi(\vec{x}_{(\zeta)}, t) \sim \frac{e^{ \frac{2 \pi}{\beta} (t_w - t) - M_{\zeta} |\vec{x}_{(\zeta)}| } }{|\vec{x}_{(\zeta)}|^{\frac{d-3}{2}} } \, .
\label{eq:ShockProfile}
\eea 
Note that the factor $2 \pi/\beta$ is the Lyapunov exponent for Einstein gravity. Note that $M_\zeta^{-1}$ defines the screening length-scale in the problem and $\lambda_L^{-1}$ defines the timescale. The butterfly velocity, as can be seen in the above equation, is a ratio of these two scales
\beq
v_B^{(\zeta)} = \frac{\lambda_L }{ M_{(\zeta)} } \quad , \quad 
M^2_{(\zeta)} =  {h_{(\zeta)}(r)}  b(r) f'(r) \sum_{\eta}  \text{dim}(\Sigma_\eta) \frac{ h'_{(\eta)}(r)}{4  h_{(\eta)}(r)} \Big\rvert_{r_0} \, .
\eeq
Here $v_B^{(\zeta)}$ is the velocity corresponding to the shockwave propagating in the $\Sigma_\zeta$ subspace. In defining $M_{(\zeta)}$ we have used the expression in Eq. \eqref{eq:KruskalMass} and switched from Kruskal coordinates to usual Schwarzschild coordinates using the identities discussed previously. For simplicity, we set $a(r) = b(r) =1$ and rewrite $M_{(\zeta)}^2$ in terms of a dimensionless quantity $\mu$, such that
\beq
\mu^2 = \frac{M^2_{(\zeta)} }{h_{(\zeta)}(r)}  =  \pi T \, \sum_{\eta}  \text{dim}(\Sigma_\eta) \frac{ h'_{(\eta)}(r)}{h_{(\eta)}(r)} \Big\rvert_{r_0} \, .
\label{eq: musqr}
\eeq

%


%

\end{document}